\documentclass[a4paper]{article}
\pdfoutput=1
\usepackage[utf8]{inputenc}
\usepackage[T1]{fontenc}
\usepackage{footnote}
\usepackage{lmodern}
\usepackage{graphicx}
\usepackage{fancyvrb}
\usepackage{multirow}
\usepackage{marginnote}
\usepackage[hyperfootnotes=false]{hyperref}
\usepackage{fullpage}
\usepackage{lscape}
\usepackage{longtable}
\usepackage{multirow}
\usepackage{url}
\usepackage{multicol}
\usepackage{tikz}
\usepackage{textcomp}
\usepackage[multiple]{footmisc}
\usepackage[final]{pdfpages} 
\usepackage{epstopdf} 
\DeclareGraphicsExtensions{.pdf,.eps,.png,.jpg,.mps}
\begin{document}
\pagestyle{empty}
\begin{center}
\begin{figure}%
\centering
\includegraphics[width=0.4\columnwidth]{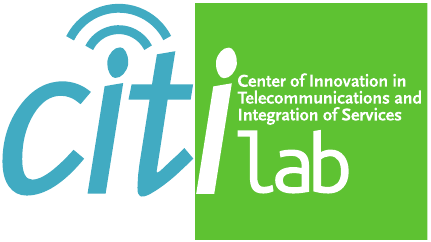}%
\end{figure}
{\LARGE \textsc{Scientific Communication}} \\
\vspace{0.5cm}
{Centre of Innovation in Telecommunications and Integration of services} \\

\vspace{3cm}
{\Large \textbf{VANET Applications: Hot Use Cases}} \\
\vspace{10pt}
{\large \textit{Marie-Ange Lèbre, Frédéric Le Mouël, \\Eric Menard, Julien Dillschneider, Richard Denis}} \\
\vspace{10pt}

\begin{minipage}{0.8\columnwidth}
\sffamily
\small
\end{minipage}
\end{center}
\vfill
\begin{center}
\includegraphics[scale=0.5]{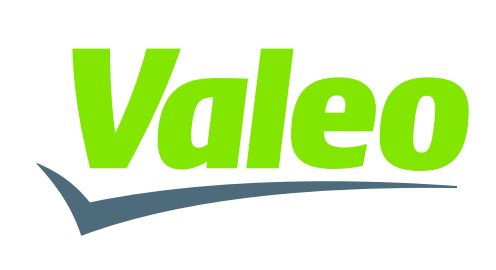}
\end{center}
\begin{minipage}[b]{0.3\columnwidth}
\includegraphics[width=\columnwidth]{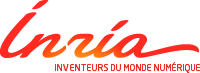}%
\end{minipage}
\hfill
\begin{minipage}{0.3\columnwidth}
\small
\centering
\textbf{
University of Lyon \\
INSA Lyon \\
INRIA\\
Valeo
}
\end{minipage}
\hfill
\begin{minipage}[b]{0.25\columnwidth}
\includegraphics[width=\columnwidth]{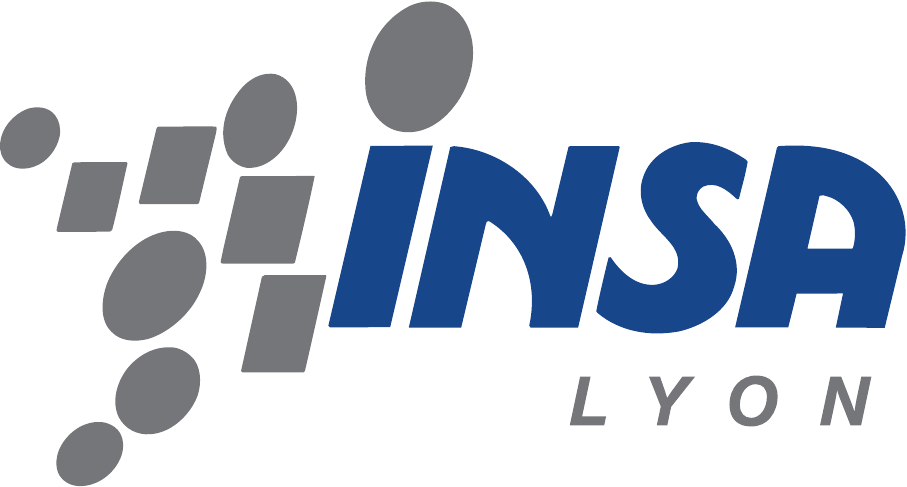}%
\end{minipage}
\newpage

\title{VANET Applications: Hot Use Cases}

\author{Marie-Ange Lebre$^{+*}$, Frédéric Le Mouël$^{+}$, Eric Ménard$^{*}$,\\
Julien Dillschneider$^{*}$, Richard Denis$^{*}$ \\[10pt]
$^{+}$University of Lyon\\
INSA-Lyon, CITI-INRIA\\
F-69621, Villeurbanne, France\\
\texttt{marie-ange.lebre@insa-lyon.fr}, \texttt{frederic.le-mouel@insa-lyon.fr} \\[10pt]
$^{*}$Valeo Interior Electronics\\
Research and Development\\
Creteil, France\\
\texttt{eric.menard@valeo.com}, \texttt{julien.dillschneider@valeo.com}, \texttt{richard.denis@valeo.com}}

\date{July 2014}

\maketitle

\newcommand{\Keywords}[1]{\par\noindent 
{\small{\em Keywords\/}: #1}}

\begin{abstract}

Current challenges of car manufacturers are to make roads safe, to achieve free flowing traffic with few congestions, and to reduce pollution by an effective fuel use. To reach these goals, many improvements are performed in-car, but more and more approaches rely on \emph{connected cars} with communication capabilities between cars, with an infrastructure, or with IoT devices. Monitoring and coordinating vehicles allow then to compute intelligent ways of transportation. Connected cars have introduced a new way of thinking cars - not only as a mean for a driver to go from A to B, but as \emph{smart cars} - a user extension like the smartphone today.

In this report, we introduce concepts and specific vocabulary in order to classify current innovations or ideas on the emerging topic of smart car. We present a graphical categorization showing this evolution in function of the societal evolution. Different perspectives are adopted: a vehicle-centric view, a vehicle-network view, and a user-centric view; described by simple and complex use-cases and illustrated by a list of emerging and current projects from the academic and industrial worlds. We identified an empty space in innovation between the user and his car: paradoxically even if they are both in interaction, they are separated through different application uses. Future challenge is to interlace social concerns of the user within an intelligent and efficient driving.\\

\Keywords{Smart Car, Vehicular Ad hoc Network (VANET), Applications, Use Case, Innovation, Vehicle to Vehicle, Vehicle to Infrastructure}
\end{abstract}

\newpage
\tableofcontents
\newpage
\section{Introduction}

\paragraph{}Vehicular ad-hoc networks (VANETs) consist of a set of vehicles travelling in urban streets and able to communicate with each others or an existing infrastructure, but without the need of a specific dedicated communication infrastructure. In fact, VANET is a special form of MANET (mobile ad-hoc network) with different nodes. Nodes are cars, buses and motorcycles moving in a restricted urban environment with constraints like traffic lights, speed limit, etc. However, unlike MANET, VANET have no problem of storage, computing power or energy saving thanks to the capacity of the car. The goal of vehicular ad-hoc network is to develop a vehicular communication system to enable quick and cost efficient distribution of data for the benefit of safety, traffic efficiency and comfort of the driver. VANET communication has recently become an increasingly popular research topic in the area of wireless networking, and takes a huge attention from government, university and industry to develop new technologies. VANET requires in its standard a need for multiple overlapping ad-hoc networks to operate with extremely high quality of service, a way to support the nature of the automotive applications (reliable broadcast) in an extreme multipath environment (reflections, high speed of the vehicles). In order to solve the problem, the Institute of Electrical and Electronics Engineers (IEEE) works on a new amendment to the IEEE 802.11 standard in order to add wireless access in vehicular environments (WAVE). They define enhancements to 802.11 which is the basis of products marketed as Wi-Fi required to support specificities of Intelligent Transportation Systems (ITS) applications. The new amendment is called 802.11p and predicts delays at the most tens of milliseconds for high-priority. Technically, a spectrum band is allocated in $5.9$ GHz for priority road safety applications, inter-vehicle communication called V2V (Vehicle to Vehicle) and infrastructure communication called V2I (Vehicle to Infrastructure). 

\paragraph{}This new way of communication between cars and infrastructure allows to imagine a lot of different applications focusing on the vehicle and on the user. In this report, we characterize the deployment and the invocation of applications for the future connected vehicle. To achieve this, we propose:
\begin{itemize}
\item a vehicular applications classification,
\item a list of recent or on-going industrial and university vehicular projects.
\end{itemize}

\paragraph{}This report is organized as follow: in the first part, we present the criteria of selection of our vehicular applications classification. Then, we list recent or on-going innovative projects and applications on the topic of smart car. We present them in three sections according to our taxonomy. Finally, we summarize all these projects with a visual classification - highlighting security and simple use cases from the automotive side, and infotainment, web-connected and social use cases from the user side. We conclude by the future convergence of these two disjointed worlds.

\section{VANET Applications: a Classification}
\label{Classification}

\subsection{Societal and Technological Criteria}

\paragraph{}To imagine the car of the future, we have to consider two important points: the vehicle it-self and its use by the applications. Our analysis classifies current innovations considering these two criteria. Two axis of complexity are defined, one is around the vehicle (horizontal), and the other is around the applications (vertical). 

\begin{figure}[!htb]
\begin{center}
\includegraphics[scale = 0.6]{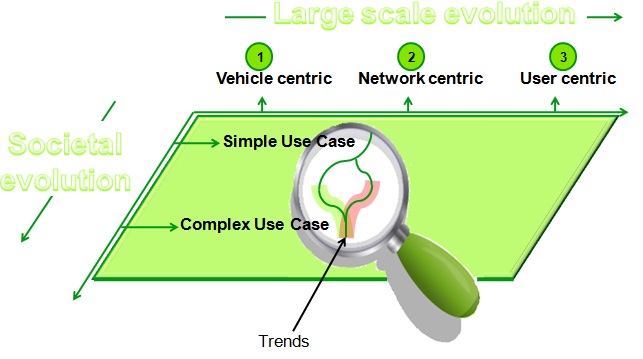}
\caption{VANET Applications Classification}
\label{ClassificationPlan}
\end{center}
\end{figure}

\paragraph{}We define the horizontal axis as considerations of the vehicle - views that can be more or less complex according to the interactions of the vehicle with its environment. First of all, a simple definition of a car can be: a fast way of transportation which allows to go from point A to point B for one or more individuals. This view is \textit{vehicle-centric}: we don't take into account other 'things' around or in the vehicle (the point 1 in the figure \ref{ClassificationPlan}). This definition is however not satisfactory as a car does not evolve alone and moves in complex environments. Whatsoever a city or a rural place, there are other elements like pedestrians, buildings; other vehicles like bus, motorcycle, taxi, truck; and an infrastructure governs its movements: intersection, traffic light, highway, etc. We fall in a \textit{network-centric} view, where all these objects are a part of a network and interact together (the point 2 in the figure \ref{ClassificationPlan}). Finally, these two approaches did not mention the most important 'things' - namely the user and the passengers in the car, we talk in this case about \textit{user-centric} view (the point 3 in the figure \ref{ClassificationPlan}). Requirements in this ultimate level are different: users have desires, behaviors, points of view, they don't have an uniform driving. For these three different points of complexity, we talk about \textit{large scale evolution} because this horizontal axis represents a graph evolution: a car (\textit{vehicle-centric}: a node) moves and communicates in a complex environment (\textit{network-centric}: a network of nodes) with its driver and passengers (\textit{user-centric}: node has social characteristics).

\paragraph{}The second vertical axis describes applications. On the previous scale, we add all the features currently inside and outside the car. Combined together, these features allow to express simple or complex applications representing the \textit{societal evolution}, the societal character and requirements of the user.\\ 

\begin{figure}[!htb]
\begin{center}
\includegraphics[scale = 0.4]{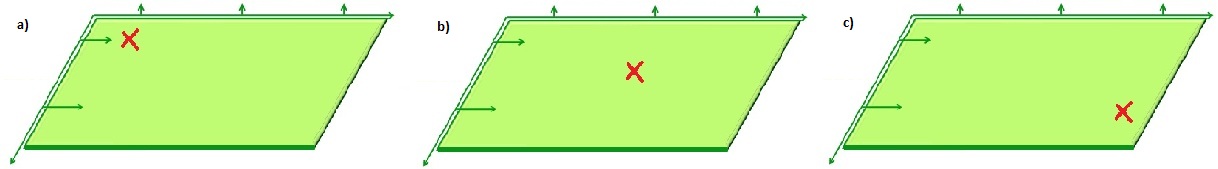}
\caption{Societal evolution: a) vehicle-centric, b) network-centric, c) user-centric}
\label{Function}
\end{center}
\end{figure}

\paragraph{}In the first level of this scale, applications warn the driver of a possible risk and give some information (\textit{vehicle-centric}, a) on the figure \ref{Function}), in a second level they are able to communicate information at the environment (\textit{network-centric}, b) on the figure \ref{Function}), and in the third level, applications take into account users with his own information, profiles and characteristics (\textit{user-centric}, c) on the figure \ref{Function}). Simple use case are classified near the coordinates ($0,0$): security, safety and traffic efficiency are currently first motivations of innovation in a vehicle. These are important concerns, so first applications deployed in the field focus on this topic. Then the more functionalities are added to the vehicle, around in the environment or provided by the user, the more the applications become complex and new trends appear. 

\subsection{Basic VANET Applications Functionalities}

\begin{figure}[!htb]
\begin{center}
\includegraphics[scale = 0.5]{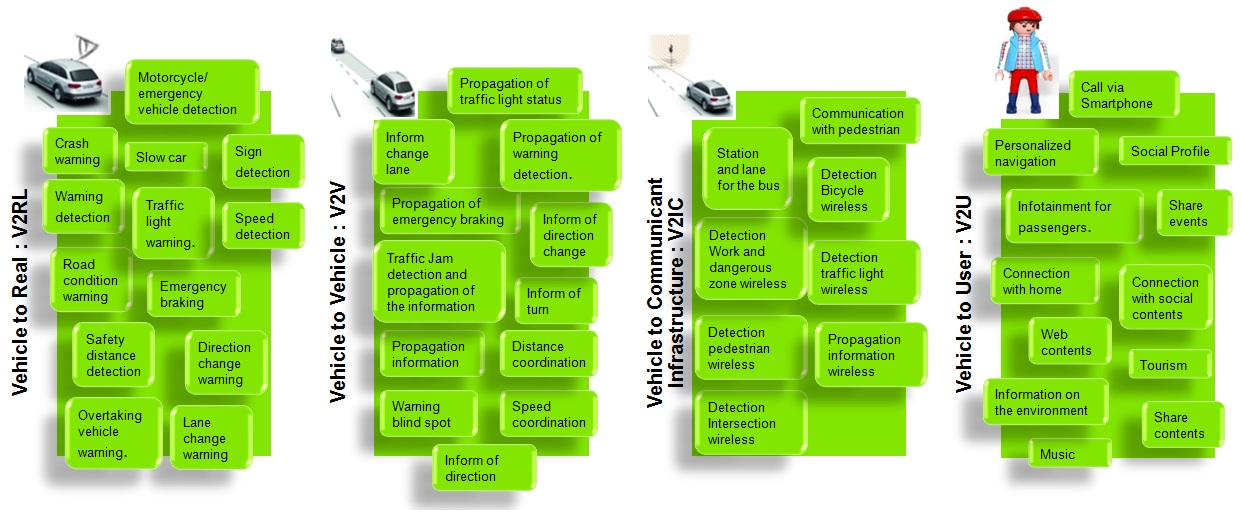}
\caption{Example of basic blocks for each view}
\label{VehicleCentric}
\end{center}
\end{figure}

\paragraph{}The complexity of a use case on the societal evolution can be considered as an addition of functionalities with software and hardware simple bricks for one vehicle. And the number of features characterizes its complexity. We define three categories to classify these basic bricks. A car interacts through its applications by different ways: the first one is to collect information from the environment, i.e. vehicle to real (V2RL). The second one is to communicate this information to other vehicles: vehicle to vehicle (V2V), and also to communicate with the infrastructure: vehicle to communicant infrastructure (V2IC) (see figure \ref{VehicleCentric}). And the latest one is to communicate with the user: vehicle to user (V2U).

\paragraph{}With only one V2RL brick, applications belong to the vehicle-centric view (the situation a) in the figure \ref{Function}). With a V2V or V2I brick, we have a simple use case in the network-centric view. If we add and connect other bricks - V2RL, V2V or V2I ones - the use case is more complex and we fall in the situation b) in the figure \ref{Function}. With V2U bricks, we integrate the user-centric view and consequently the use case presents social characteristic. 

\subsection{Scenario Examples}

\subsubsection{Network-centric view}

\begin{figure}[!htb]
\begin{center}
\includegraphics[scale = 0.5]{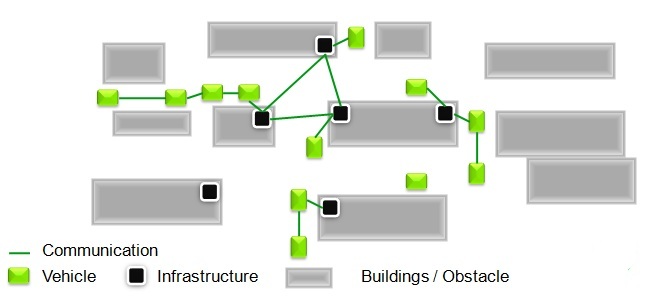}
\caption{Network-centric example: Urban V2V and V2I applications}
\label{NetworkCentric}
\end{center}
\end{figure}

\paragraph{}In the network-centric view, the car has the possibility to communicate with its environment: traffic light, buildings, vehicles, pedestrians, etc. So we consider a network with several vehicles and with different infrastructures, and, consequently, several different parameters need to be taken into account such as the mobility, the wireless communication, the building interferences, etc (see figure \ref{NetworkCentric}). 
\subsubsection{User-centric view}

\paragraph{}In the next level of complexity - \textit{user-centric} view - we take into account drivers with their relationships, desires, personalities, etc. To build such use cases, applications can be build by directly integrating V2RL, V2I and V2U blocks, but many projects derive from vehicle-centric and/or network-centric existing applications and just add V2U blocks to these applications. In this case, use cases become even more complex to preserve the backward compatibility.

\begin{figure}[!htb]
\begin{center}
\includegraphics[scale=0.9]{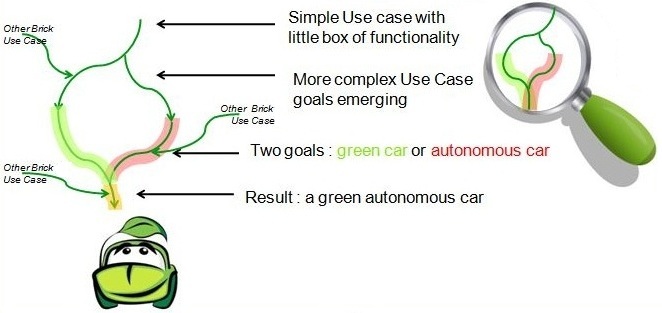}
\caption{User-centric example: Green, Autonomous Cars}
\label{ApplicationEvolution}
\end{center}
\end{figure}

\paragraph{}In the actual context of automotive, applications are deployed for security, safety, traffic efficiency, energy saving, therefore, their user-centric evolutions include infotainment for passengers, user-awareness guidance and so vehicle tends to be more autonomous (smart) and ecologic (green). The figure \ref{ApplicationEvolution} illustrates this process of adding bricks which will give particular social characteristics to the vehicle of the future.

\section{Emerging Trends in SmartCar}

\paragraph{}First section presents the most mature V2RL innovative projects - i.e. all current applications where the vehicle is able to catch its environment, for instance, driving assistant with the use of technologies like radar, camera, laser. Second section details V2V and V2I applications - together as many projects present them in a complementary way. Finally, we focus on V2U applications with existing and futuristic projects.

\subsection{V2RL}

\subsubsection{Honda: Traffic Congestion Minimizer}
\marginpar{\includegraphics[scale = 0.05]{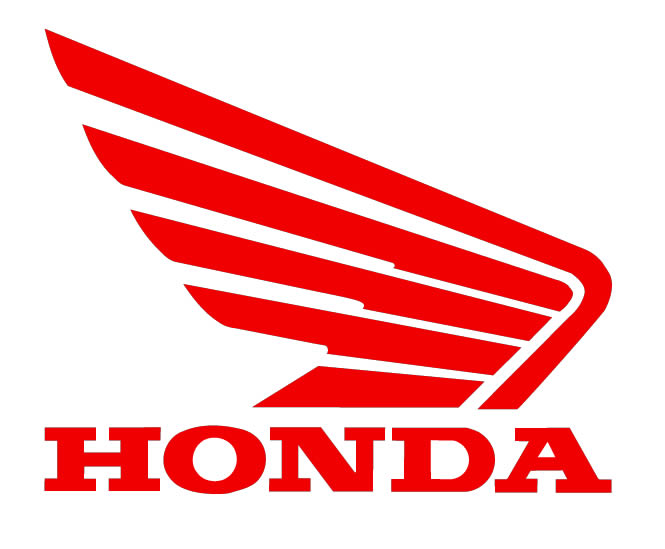}}

Honda's project \cite{HondaTraffic} was presented in Tokyo, 26 April 2012. Its goal is to reduce traffic jam, especially when traffic is made worse by the accordion effect - sudden braking and rapid acceleration. With the use of accelerometers, the proposed system measures the acceleration and deceleration of the vehicle. Then, thanks to a red/green color code displayed on the dash, the system indicates if the driving is too erratic and gives consequently advice for a better driving. One car equipped with this system in a traffic jam is however useless. The system proposed links so multiple cars via a web-connected cloud server. The other drivers are, therefore, aware of the driving patterns of vehicles ahead and the cruise control system is so able to maintain a constant distance between vehicles. A public-road testing has begun in Italy in May 2012 and Indonesia in July 2012 to analyze its efficiency.
\begin{figure}[!htb]
\begin{center}
\includegraphics[scale = 0.6]{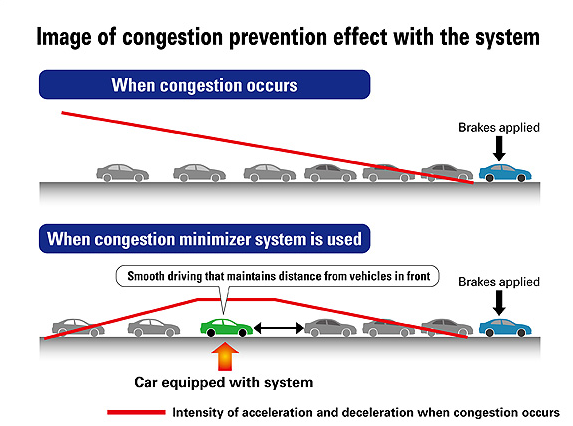}
\caption{Accordion effect prevention}
\end{center}
\end{figure}

\subsubsection{Pioneer: 'Cyber Navi' or 'AVIC-ZH99HUD'}
\marginpar{\includegraphics[scale = 0.2]{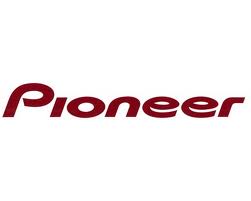}}

Pioneer presented their product \cite{Pioneer} for the first time in May 2012 in Tokyo and got it to the market in July 2012 - only in Japan. 'Cyber Navi' is a GPS in augmented reality. 70\% of cars are compatible with the system. Equipments are a Head's Up Display - a laser projector installed in the place of the sun visor; a camera fixed on the rear view central mirror and an LCD screen completing the system. An image measuring 90x30cm is projected 3m in front of the driver. Information displayed are speed, way identification, security distance, speed limit, traffic light and GPS coordinates and news. 

\subsubsection{BMW: Head-Up Display (by Continental)}
\marginnote{\includegraphics[scale = 0.2]{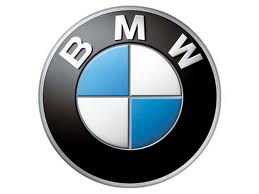}}
\label{BMWAR}

BMW, in partnership with Continental, focuses on reality augmented for navigation assistant with an Head-Up Display \cite{BMWHUD}. A first generation was marketed in 2004 with a display on the windshield of the navigation directions, speed limits, lane departure warning, night vision alerts and check/control messages. Currently, Continental is providing Head-Up Display equipments for the new BMW 3 Series.
\begin{figure}[ht!]
\begin{center}
\includegraphics[scale = 0.5]{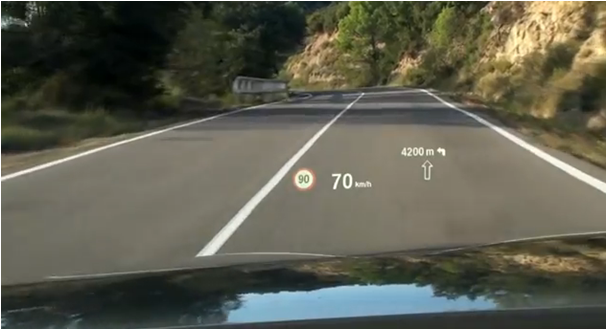}
\caption{First generation of reality augmented by Head-Up Display}
\end{center}
\end{figure}

\paragraph{}The study of the next generation displays - tested in simulators - began in 2011 \cite{BMWHUD2} with virtual markers superimposed on the real world: lane boundaries and optimal route based on digital road map information.
\begin{figure}[!htb]
\begin{center}
\includegraphics[scale = 0.5]{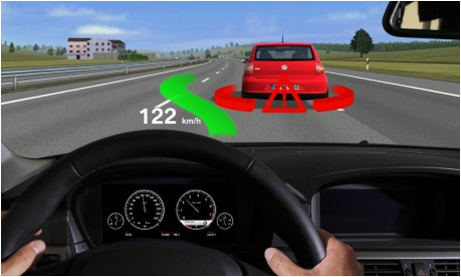}
\caption{Future generation of reality augmented by Head-Up Display}
\end{center}
\end{figure}

\subsubsection{Intel: Augmented Reality and In-vehicle Infotainment System}
\marginpar{\includegraphics[scale = 0.2]{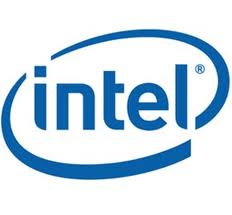}}

Yi Wu, a senior research scientist in interaction and experience research at Intel Labs in Santa Clara, works on \cite{Intel}: 'Augmented Car Navigation System' (Windows and Windshield display information for navigation). Intel also has collaboration for work on 'in-vehicle infotainment system' in order to deliver information (on the navigation) and entertainment (music, video, radio, etc.) to drivers and passengers (via touch screen, voice recognition, gesture). They work with Nissan, Kia Motors, Toyota, Denso, China TSP, GAIG Guangzhou Automobile Group and Hawtai motor, and focus on fusion between mobile and vehicles, cloud-based services, control vehicle via smartphone (virtual key, video surveillance) and connectivity (internet, 3G for email, web, mobile function, TV). On their research user will be able to catch important information on the web as the best price of a gas station or the nearest coffee, he also will be able to pay and download music, pay professional bills, send e-mails, read them, etc. They also proposed safety applications like 'keeping your lane'. 

\subsubsection{PSA Peugeot Citro\"en: 'Peugeot connect Apps'}
\marginpar{\includegraphics[scale = 0.4]{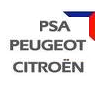}}

PSA Peugeot Citro\"en proposed in October 2012 \cite{PSA} a touchscreen interface with essential Web applications in the vehicle: traffic information, navigation information or how to find the nearest parking available everywhere in Europe. The maintenance of the car can be checked on smartphone via an 'itouch' application. 

\subsubsection{BMW: 'XL Journey Mate Mini'}
\marginpar{\includegraphics[scale = 0.1]{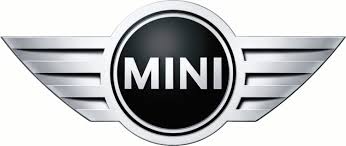}}

This application marketed on Mini in 2013-2014 helps for navigation with real traffic information. The system knows the driver's preferred route, reminds him when it is time to refuel, messages him about impending appointments or planned telephone calls, reports traffic information in real time, assists in locating parking spaces, and directs the driver on foot to his final destination. The application is also able to remind the user of calendar entries in his smartphone. 

\subsubsection{Land Rover: 'Concept Discovery Vision'}
\marginpar{\includegraphics[scale = 0.2]{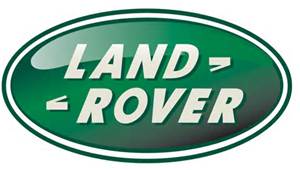}}

This concept car uses infrared lasers (placed in the fog light) to scan the ground in front of the vehicle. These sensors anticipate the reaction of the vehicle. They allow also to estimate the depth of a stream or pond before the car enters. The driver can control his car at distance in order to overcome obstacles.  The car will handle the crossing. A semi-autonomous driving at low speed is possible. User can project images on the road with the laser light (triangle). This concept proposes also to control functions with gestures and voices. Car cameras can also project video on smart windows. Finally for helping the driver, an HUD can offer a digitized view of the ground on the hood, therefore, the front of the vehicle is virtually invisible from the cockpit.

\subsubsection{SystemX: Localisation in Augmented Reality}
\marginpar{\includegraphics[scale = 0.2]{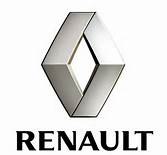}}

This project which begins in 2013-14 is a reflection on the autonomous vehicle, with an approach that involves studying new interactions with the driver in 2020. LRA project will examine more specifically the delegation of driving on freeway or expressway. In several cases (traffic flowing traffic), engineers will validate, with the help of monitoring software of the pipe (which will promote a range of partial and targeted behavior), the most appropriate interfaces. Different partners are on this project: Research Technology Institute (IRT), Alstom Transport, M3 Systems, Oktal, Renault, Safe-River, SysNav, Valeo and CEA-List.
\begin{figure}[!htb]
\begin{center}
\includegraphics[scale = 0.3]{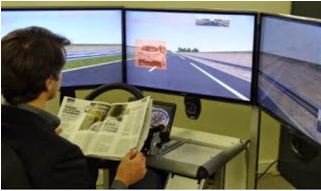}
\caption{French Project 2014 on augmented reality}
\end{center}
\end{figure}

\subsubsection{Siemens: Navigation system in Augmented Reality}
\marginpar{\includegraphics[scale = 0.1]{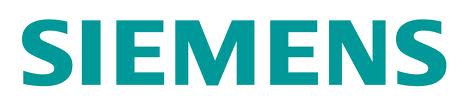}}

The University of Linz and Siemens \cite{Siemens} worked on a navigation system based on a video technique: a see-trough HMD (Head-mounted display) with hand-held display. This project started in 2000 and finished in 2005.
The devices are a GPS, a map of the area (topography), a computer software, a camera on the windshield, an odometer to measure speed, a gyroscope to measure or maintain orientation, a choice of several view: bird's eye view on the highway and augmented reality near intersections.\\
It was a concept, and the next step of their study is to project the information on the windshield.
\begin{figure}[!htb]
\begin{center}
\includegraphics[scale = 0.3]{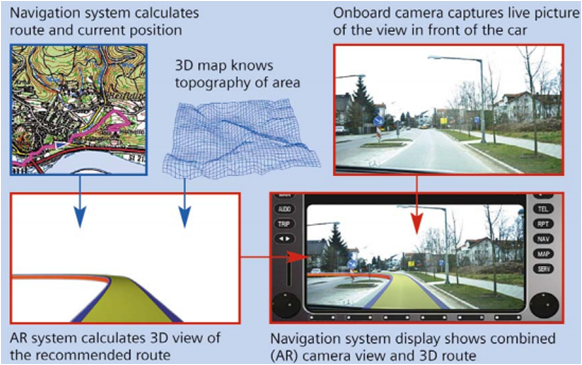}
\caption{Navigation system of Siemens and University of Linz}
\end{center}
\end{figure}

\subsubsection{Microsoft: iOnRoad}
\marginpar{\includegraphics[scale = 0.2]{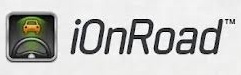}}

Lot of smartphone applications are similar, we choose only the most popular and the highest rated. This smartphone application exists since 2010. Reality augmented (with the camera of the smartphone) allows audio and visual alerts for the safety distance. The originality is the other functions of the application: the user can accumulate points of security with keeping safety distance during travels and share his result with other drivers on his web dashboard ('Driving Profile') and also on Facebook. He can a share wonderful view and all his snapshots. Naturally this application includes classic phone apps available with a click on the screen. The system can read and notice that user has received an SMS, and finally when he parks his car, iOnRoad keeps the GPS location and takes a photo in order to find the car easily later.
\begin{figure}[!htb]
\begin{center}
\includegraphics[scale = 0.4]{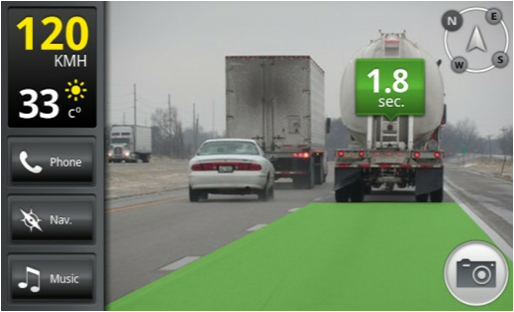}
\caption{Screen of the smartphone with the iOnRoad}
\end{center}
\end{figure}

\paragraph{}However, some problems are not fixed: Safety distance in traffic jam are always in red, and the application is unavailable during the night.

\subsubsection{Google: Driverless Car}
\marginpar{\includegraphics[scale = 0.1]{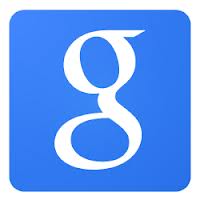}}

Google \cite{Google} has a project around a car which is able to go from point A to point B without driver. The vehicle is equipped with a Laser LIDAR (Light Detection And Ranging Light) which is an optical remote sensing technology (create a 3D image). It captures moving object, other cars, pedestrians, cyclists, stationary objects, road signs and stop lights. The car also has radar sensors and a position sensor attached to one of the rear wheels that help locate the car's position on the map. An algorithm manages changing lane and the system uses information from Google Street View. Google's car yields the right of way according to the traffic laws. At the moment Google's cars are allowed to drive in Nevada (since June 2011) and Florida and California (since September 2012). The disadvantage is the price of the equipment: 100 000 euros for one car.
\begin{figure}[!htb]
\begin{center}
\includegraphics[scale = 0.4]{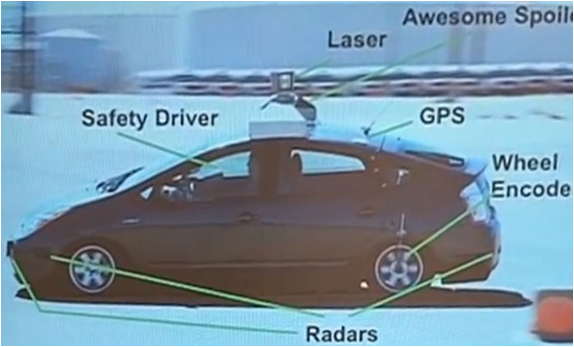}
\caption{Driverless Car Google}
\end{center}
\end{figure}

\subsubsection{French Automatisation Basse Vitesse (ABV) project}
\marginnote{\includegraphics[scale = 0.2]{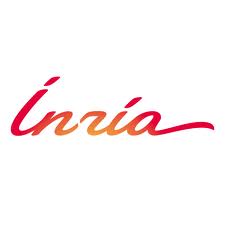}}

\begin{figure}[!htb]
\begin{center}
\includegraphics[scale = 0.6]{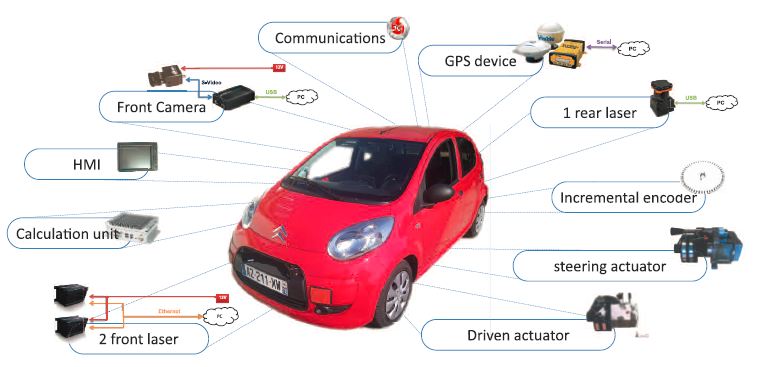}
\caption{ABV Project}
\end{center}
\end{figure}
The project wants to demonstrate the technical feasibility of fully automated driving at speeds below 50 km/h like the previous project by Google. Researchers of Inria are in charge of the automation of an electrified Citr\"{o}en C1. ABV system relieves the human driver from performing monotonous tasks such as holding the brake pedal or rather risky maneuvers like changing lanes or keeping safe distance from the vehicle in front. When the application finishes its task the driver has to take the control of the vehicle, if he fails to do so, the vehicle will automatically stop. Algorithms follow the classical scheme: perception, planning and control. They test the Citr\"{o}en on the Satory tracks (Versailles, France). The vehicle is equipped with sensors (odometers, lasers, frontal camera) in order to detect obstacles, to avoid and determine which lane to follow. It has a computer to process data, make decisions like acceleration/brake. It also has steering actuators to control the vehicle and a communications devices.

\subsubsection{Volvo : Magnet Project}
\marginnote{\includegraphics[scale = 0.2]{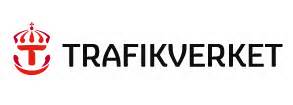}}

Volvo has deployed 100 vehicles in Gothenburg in march 2014, they are testing a technology that could be insert into traffic with facility. Magnets are embedded in the pavement, they were implanted under the ground at a depth of 200 mm. A solution that is simple, effective and inexpensive to improve safety. The test car is equipped with a lot of magnetic field detectors. The brand believes that, contrary to established positioning technologies such as GPS and cameras that admit their limitations in certain conditions, magnets embedded in the pavement are not sensitive to physical obstacles and weather conditions. This is a partnership with the Swedish Transport Administration.
\begin{figure}[!htb]
\begin{center}
\includegraphics[scale = 0.6]{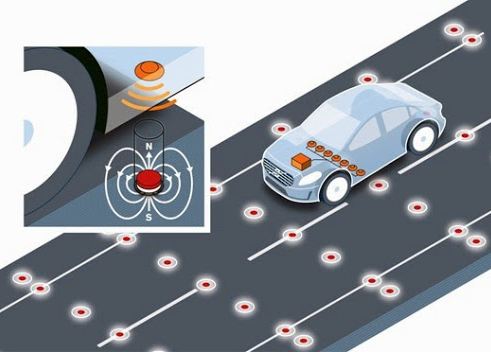}
\caption{Magnet Project}
\end{center}
\end{figure}

\subsubsection{Ford : 'Blueprint for Mobility'}
\marginnote{\includegraphics[scale = 0.2]{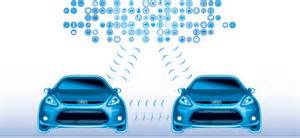}}

In the Mobile World Congress 2014 Ford present its project on the autonomous car. The vehicle will have 4 LIDAR sensors that perform 2.5 million measurements per second and can detect other vehicles, cyclists, pedestrians and even small animals. The car has a 3D map in real time which allow it to see what is happening around it at a distance of 70 m. Ford initiates a partnership with the Technical University of Aix-La-Chapelle to develop man-machine interfaces compatible with the future autonomous cars. And they also take as partnerships : the MIT (Massachusetts Institute of Technology) and the University of Stanford. Ford and MIT seek to develop learning algorithm for anticipate the movement of vehicles and pedestrians. Ford and Stanford search to analyze the area behind an obstacle. Their goal is to equip vehicles with a capacity of judgment. 

\newpage

\subsubsection{Move'o : Autonomous Car}
\marginnote{\includegraphics[scale = 0.2]{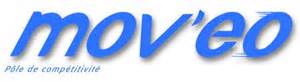}}

The competitiveness cluster R\&D Automotive and Public Transportation Move'o work on innovative projects to strengthen the international competitiveness of french companies and territories. In 2014 Mov'eo decide to form a french group of 'PME' specialized in high technology in order to provide a comprehensive response to industrial group in the context of the driver assistance systems. Seven PME include all skills and technologies to meet current and future needs of automotive manufacturers and suppliers.

\subsubsection{DriverlessCar: other projects}
\marginpar{\includegraphics[scale = 0.7]{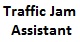}}

BMW \cite{BMWTrafficJamAssist} works on a similar project based on semi-autonomous driving system on highway and for traffic jam assistant. Audi \cite{AudiTrafficJamAssist} has a project of research on a traffic jam assistant too (January 2012). Ford \cite{FordTrafficJamAssist} announced in June 2012 the development of the 'Traffic Jam Assist' for help relieve road congestion and relaxing the driver. Volkswagen \cite{VolkswagenTrafficJamAssist} presented a project in June 2011 (a use case demo) which is a temporary auto-pilot. Mercedes works also on traffic jam assist (for 2013). General motor \cite{GMTrafficJamAssist} works on it since 2008. Cadillac \cite{CadillacTrafficJamAssist} in 2012 presents 'Super Cruise' (just on Highway, and the detection of lane departure when it is not clear is a real problem). Inria (Institut National de Recherche en Informatique) works on the \textit{Cybercar}, the car is view like a horizontal lift with the concept of collective vehicle (cars and Bus without driver). With optic and magnetic sensors, cameras and lasers the car will be able to transport an individual from a point A to a point B.

\subsection{V2V and V2I}
\paragraph{}In this section we will present an overview of the most popular ongoing projects around V2X communication in the world, and also some finished projects but innovative (See \cite{Planning2012} for a detailed survey of projects in each country around the world). We decided to aggregate V2V and V2I applications because they both focus on the security and the traffic control. They are complementary because an application will use generally bricks in V2V and in V2I together.

\subsubsection{Intelligent Truck Parking Application}
\marginnote{\includegraphics[scale = 0.2]{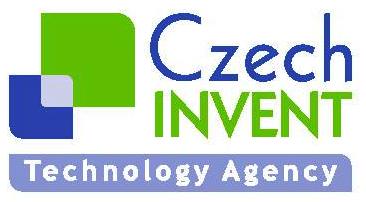}}

The first example of cooperative system with V2V and V2I is the Intelligent Truck Parking Application (ITP), an R\&D project funded by the Technology Agency of the Czech Republic. The goal is to predict the occupation of truck parking places on highways and immediately inform drivers via V2X technology to help them plan their rest stops in accordance with legislation. Moreover historical data from the tolling system is incorporated with actual data from traffic sensors (loops, video, detectors). The next step is to validate outputs of the model and calibrate it to improve its accuracy.

\subsubsection{Vision of Internet of Cars}
\marginnote{\includegraphics[scale = 0.15]{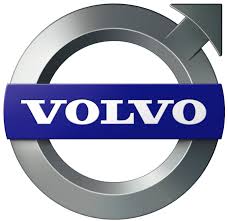}}

Volvo and SICS Swedish ICT have launched a collaborative project to open the computer systems of cars for the market of applications. They increase their effort to work together on the future vision of the internet of Cars (IoC) with embedded systems that interact to provide new services to their users. The idea is to let vehicles exchange information that they normally collect like the state of the road, traffic jams, the exact time of arrival, etc. This information can contribute to making trips safer, however this requires an entirely new level of security and robustness. In this project a framework for opening up computer systems of vehicles is being developed. The simulation work will be transferred from a PC to embedded hardware. A couple of credit-card computers, Raspberry Pis, will be interconnected and used to simulate electronic control units of a real car. An example of an application: if the speed information is exchanged directly between cars, the car can itself detect a stop in the traffic ahead and adapt its speed to avoid a sudden break. Currently they conduct an empirical research based on case studies in order to try to identify the primary interfaces between stakeholders in a typical IoC-system and a first demonstration is expected at the end of 2014.

\subsubsection{An autonomous traffic warning system with Car-to-X communication}
\marginnote{\includegraphics[scale = 0.15]{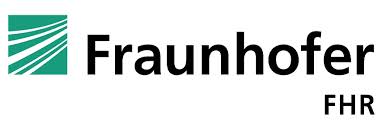}}

Researchers of Fraunhofer work on a system which uses a combination of stationary sensors like radars, loops and information relayed from vehicles. The system analyzes the data and identifies dangerous situations like sudden appearance of road construction sites, traffic back-ups, object on the road or bad weather. By this system drivers are informed of the precise location of the danger through warning lights integrated in the reflectors posts and directly via car to x communication. Matching the information from multiple sensors covering the same area ensures a high level of data quality at the source. The features used in their system are: ITS G5 and GeoNetworking, CAN bus, interfaces to external sensors, digital map, CAM and DENM management, precise determination of the position of the vehicle and connection with a HMI like touch screens or tablets. The concepts and processes are validated in a simulation environment and then implemented with in-house developed software and hardware on a test route. The project is funded by the Bavarian Ministry of Economic Affairs, Infrastructure, Transport and Technology.

\subsubsection{Volvo Car: SARTRE}
\marginnote{\includegraphics[scale = 0.2]{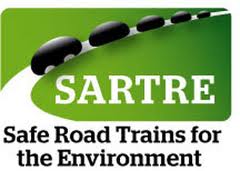}}

This project \cite{SARTRE} was launched in September 2009 to September 2012 by the European Commission under the Framework 7 program. Participants are Idiada and Robotiker-Tecnalia of Spain, Institute Kraftfahrwesen Aachen (IKA) of Germany, Technical Research Institute of Sweden, Volvo Car Corporation and Volvo Technology of Sweden (the only car manufacturer). The principle is a Safe Road Trains for The Environment (SARTRE) with the formation of platoons on highways. A professional driver will take responsibility for platoon and drivers can do other things in the car in a safety environment. Technologies used in this project are camera, radar, laser technology (they support collision detection, adaptive cruise control, lane keeping aid, blind spot information, park assist), an HMI (Human Machine Interface) with a touch screen for displaying vital information (possibility to leave or join the platoon) and V2V technology with 802.11p to control and coordinate the movement of the platoon. Moreover V2I technology provides services for making platooning economically feasible, and guiding vehicle to suitable platoon. Advantages are safe transport (professional driver), environmental impact reduced and improvement of the traffic flow (reduced speed: 90 km/h on highway). Their future study is based on the control of an emergency situation like obstacle avoidance or sudden braking. Some functionalities will be available in 2014 for the traffic jam (until a speed of 50 km/h).
\begin{figure}[!htb]
\begin{center}
\includegraphics[scale = 0.6]{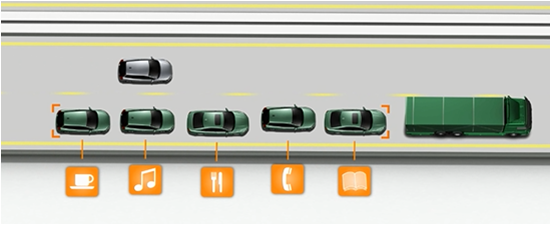}
\caption{SARTRE Project}
\end{center}
\end{figure}

\subsubsection{Audi: 'Travolution project'}
\marginpar{\includegraphics[scale = 0.3]{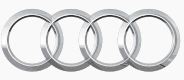}}

Audi \cite{AudiTrav} started this project in 2006 and latest results were presented in Ingolstadt in June 2010. It was based on the communication with the traffic light using wireless LAN and UMTS (Universal Mobile Telecommunications System) links. The human machine interface displays a recommended speed when the vehicle is approaching a traffic light for avoiding to stop, in this case the adaptive cruise control (ACC) is ready, then speed was accepted and controlled via the ACC.\\
If the car is stopped at red light, the driver knows how long he has to wait before the light switches to green again. They predict a greater traffic flow and a lower CO2 emissions with 17\% of fuel savings and 15\% reduction in CO2 from urban traffic. They also study payment online when parking or refueling the car with a communication with the stationary equipment.

\subsubsection{Swarco}
\marginpar{\includegraphics[scale = 0.3]{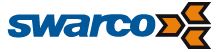}}

Swarco traffic system works since 2010 on cooperative mobility benefits and traffic light optimization. They are based in Austria. They maintain equipment at intersection and also install traffic controllers. They are member of the Car2Car communication consortium (it consists of all relevant automotive OEMs and various suppliers). In 2012 they test different use cases with 802.11p on six cooperative traffic controllers in a city: red light violation warning, remaining phase duration, green wave speed advisory and GLOSA (Green Light Optimal Speed Advisory). They use different messages which are in the European architecture of 802.11p: SPAT (Signal Phase and Timing Message) in order to know traffic light timing, DENM (Decentralized Environmental Notification Message) to warn approaching vehicle and CAM (Cooperative Awareness Message) for emergency vehicle that requests a prior green from the traffic controller. In the future they want to study public transport prioritization, accident aware traffic, intelligent signal heads for displaying remaining waiting time and also controller based only on DENM messages.

\subsubsection{Volkswagen: 'C3World, connected car in a connected world'}
\marginpar{\includegraphics[scale = 0.1]{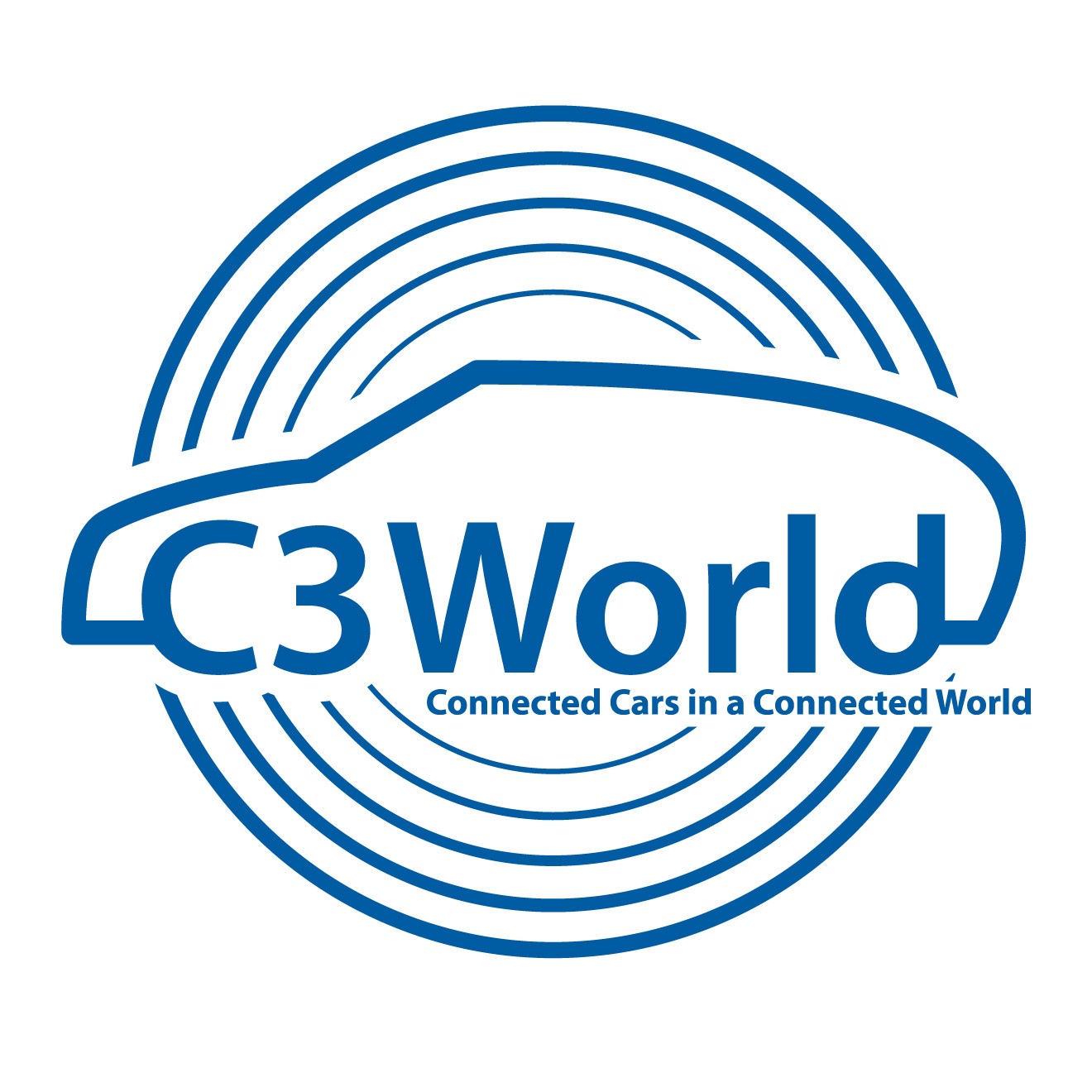}}

This project \cite{C3World} started in 2007 and latest results were presented at the CeBIT 2012. This consortium was created by Volkswagen and three research institutes in order to conduct research into information and communication technology for the vehicle of tomorrow. Research partners are the Institute for Communications Technology of the TU Braunschweig, the Institute for Communications Technology of the Leibniz University in Hannover and the OFFIS Institute for Information Technology in Oldenburg. Research focus on robust and secure wireless transmission of information within the vehicle and between the vehicle and its environment. There are three research topics. First one is car to X communication which includes applications for the security, simulations of these application, studies of a 'hybrid network' (broadcast services, IEEE 802.11 WLAN networks), traffic simulators and environment models. The second one is equipment and data links in the vehicle which focus on coupling mobile devices with vehicle and UWB interfaces in combination with the approach employed by an open software platform. The latest is a location-based web search in the vehicle for a suitable user interface, an intuitive access and a system for filter the data from the internet according to context and driving situation.

\subsubsection{Denso Corporation: Field Operational Test}
\marginpar{\includegraphics[scale = 0.05]{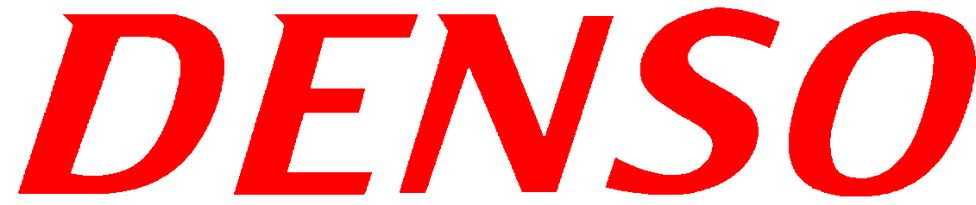}}

Denso collaborated \cite{Denso} with Tongji University in Shanghai and began a V2X technology field test in March 2012 on public roads (Denso also supports the Connected Vehicle Safety Pilot Program in August 2012 in USA (see section \ref{SafetyPilot})). Their tests are based on position and speed of emergency vehicles (ambulances and fire engines) which are communicated to the surrounding vehicles and roadside infrastructure. When an emergency vehicle is approaching, traffic light changes at intersections and alert surrounding vehicles to switch lanes. DENSO works on V2X technology since 2003. 

\subsubsection{SmartWay}
\marginnote{\includegraphics[scale = 0.7]{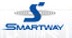}}

SmartWay \cite{SmartWay} was an autonomous systems field operational test from 2004 to 2010 in Japan. It was Conducted by the National Institute for Land and Infrastructure Management (NILM) with 40 vehicles and 2 522 test runs in 7 different towns and 23 private partners. They allow to develop VICS (vehicle information and communication system technology in japan) and ETC (Electronic Toll Communication). VICS provides traffic and travels information in real time on a human machine interface in the vehicle. Currently 1600 ITS spots are deployed in Japan (for VICS) and 5,8GHz is use for ETC. 
\begin{figure}[!htb]
\begin{center}
\includegraphics[scale = 0.7]{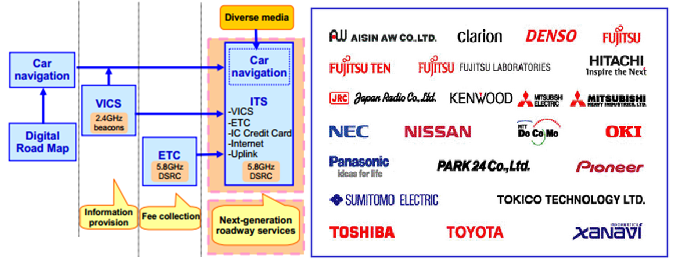}
\caption{SmartWay function and Partners}
\end{center}
\end{figure}

\subsubsection{US Department of transportation: Connected Vehicle Safety Pilot Program}
\label{SafetyPilot}
\marginpar{\includegraphics[scale = 0.7]{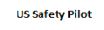}}

The test started in August 2012 for one year \cite{USSafetyPilot}. The aim is to collect empirical data to present a more accurate detailed understanding of the potential safety benefits of these technologies. It is conducted by Research and Innovative Technology Administration (RITA). And participants are Ford, General Motors, Honda, Hyundai/Kia, Mercedes-Benz, Nissan, Toyota and Volkswagen/Audi, for all 2,800 cars, trucks and buses. For equipment, Kapsch, ITRI International Inc, Cohda Wireless, Cisco Systems Inc and Savari networks were chosen.\\

\paragraph{}The test takes place in Ann Arbor, Michigan with 29 Roadside units and 73 lanes miles of roadway. There are three kinds of devices. The first one is embedded devices, installed during the manufacturing process (integrate with the vehicle's computers) thus providing the ability to draw on a wide range of data. They emit and receive the basic safety message, communicate data on speed, acceleration and deceleration, yaw rate, turning, wiper activity, and braking, among others. The second one is aftermarket devices, they don't connect to the vehicle's computers, they draw data only from the environment (GPS, safety messages from other vehicles) to support applications. They can emit the basic safety message to warn equipped vehicles of their presence as well as warn drivers of potential conflicts. The last one is vehicle communication devices which focus only on emitting the basic safety message (BSM).\\

\paragraph{}All of these devices emit a BSM 10 times per second, which forms the basic data stream that other in-vehicle devices use to determine when a potential conflict exists. The key applications tested include forward collision warning, emergency electronic brake light, intersection movement assist, blind spot, lane change warning, don't pass warning and curve speed warning. They use DSRC technology which can provide sufficiency low latency (100 ms) to be suitable to prevent crashes associated with collision at a blind intersection, vehicle changing lanes in blind spot and vehicle stopped ahead.\\
The challenge is at the end of the program in 2013, namely NHTSA (National Highway Traffic Safety Administration) agency will take a decision on how and when to deploy V2V and V2I systems.

\subsubsection{Compass4D}
\label{Compass4D}
\marginpar{\includegraphics[scale = 0.3]{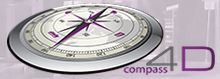}}

This is an European project coordinated by ERTICO (This is a network of Intelligent Transport Systems and Services stakeholders in Europe, they connect public authorities, industry players, infrastructure operators, users, national ITS associations and other organizations together). Compass4D is a cooperative mobility pilot on safety and sustainability services for deployment, it began in January 2013 and will finish in December 2015. They study three services: forward collision warning, red light violation warning and energy efficient intersection service. Their goals are to prove the positive cost benefit, lead organizations to an international cooperation and standardization and obviously ensure the deployment of their three services. Seven European cities are included in this project and thirty one partners in Europe.

\subsubsection{'Safe Intelligent Mobility-Test Field in Germany' SimTD}
\label{SIMTD}
\marginpar{\includegraphics[scale = 0.3]{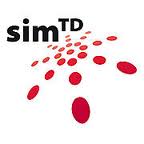}}

This project \cite{SimTD} began in 2008 and lasted 4 years. The field operational test which is the latest part of the project has begun in August 2012. The aim is to evaluate the benefits of V2X communication for traffic, safety and efficiency and also study user acceptance. Participants are Audi, BMW, Daimler, Ford, Opel, Volkswagen, in all 120 cars equipped drive in Frankfurt. Suppliers are Bosch and Continental, telecommunication and services provider is Deutsche Telecom, main research institutions and universities are Fraunhofer, University Wurzburg, Munchen and Berlin, DFKI (German Research Centre for Artificial Intelligence) and finally several public authorities are associated with Car 2 Car Communication Consortium \cite{Strang}. The devices used are 100 Intelligent Roadside Stations (IRS), two ITS central stations (ICS), 120 ITS Vehicle Station (IVS) with an application unit (AU) and a Communication Control Unit (CCU) based on 802.11p, UMTS, IEEE 802.11b and IEEE 802.11g (WLAN access). CCU has a DGPS (Differential Global Positioning System is an enhancement to GPS that provides improved location accuracy) and is connected to the AU via Ethernet, AU is based on JAVA/OSGi.
\begin{figure}[!htb]
\begin{center}
\includegraphics[scale = 0.4]{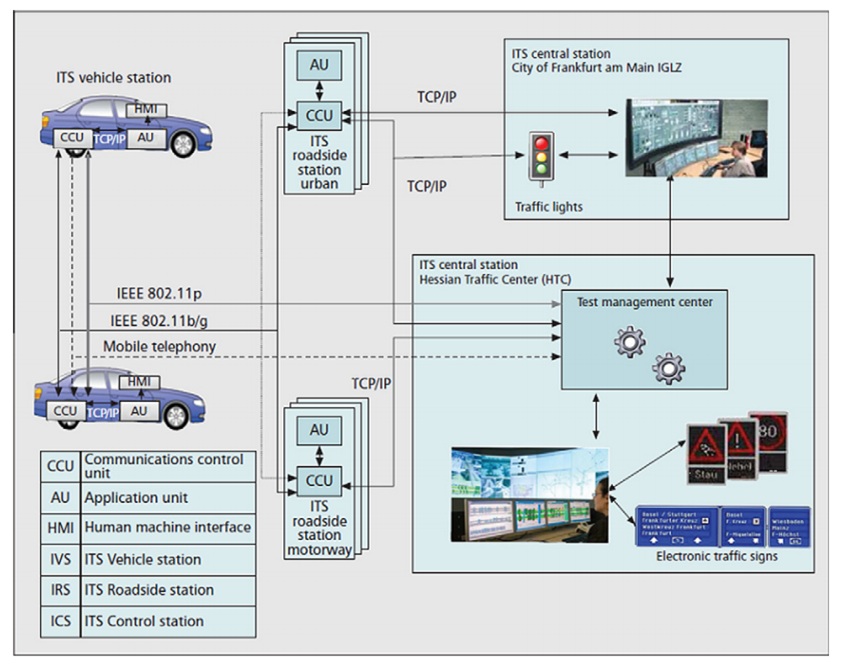}
\caption{SimTD equipment}
\end{center}
\end{figure}

\paragraph{}Current V2X research projects study only a small number of applications, and the system is optimized for these applications.
SimTD has the exact opposite approach, instead of optimizing the system they try to cover all three main categories: safety, traffic efficiency and commercial services. Thirty four use cases in all these three categories are tested.
\begin{figure}[!htb]
\begin{center}
\includegraphics[scale = 0.4]{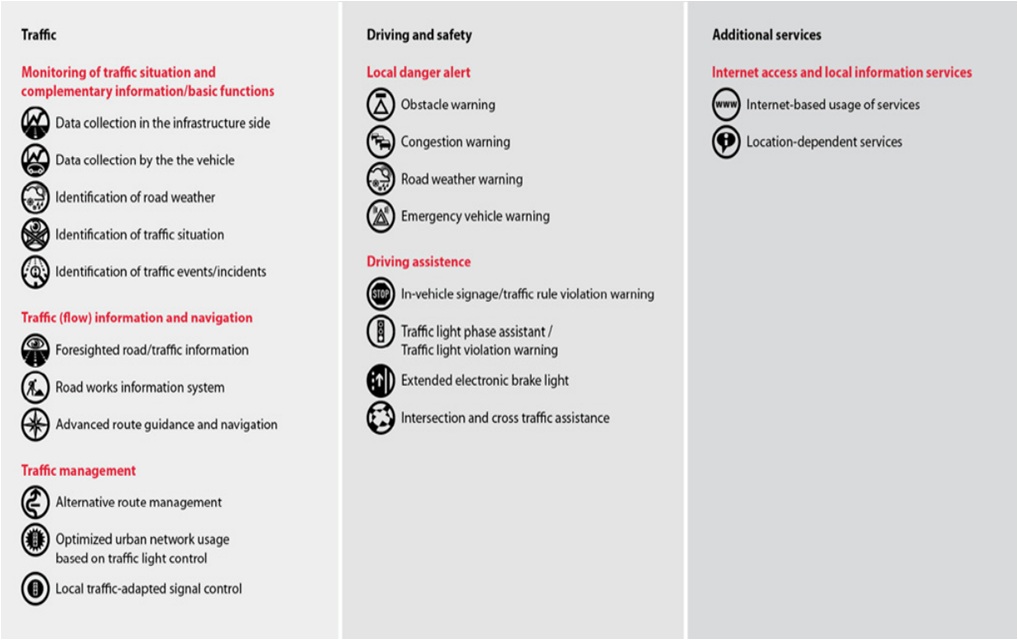}
\caption{SimTD project applications}
\end{center}
\end{figure}

\subsubsection{SCORE@F in France}
\label{SCORE@F}
\marginpar{\includegraphics[scale = 0.1]{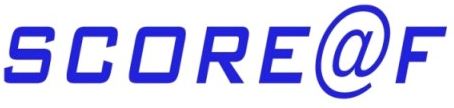}}

The project is similar to SimTD but tested in France. SCORE@F (System Cooperatif Routier Experimental Francais, 2010-2013) led by Renault contains 12 industry partners and 7 laboratories. The goal is to quantify benefits of V2X, identify stakeholders, validate standards and develop applications. The community will use 30 vehicles in order to test among others, use cases on cooperative awareness, longitudinal risk warning, intersection collision risk warning, traffic light violation warning, green light optimal speed advisory and electric vehicle charging.
PSA and Renault are involve in this project and the next step (-2016) is to deploy 2000 vehicles for test in real conditions. 

\subsubsection{Other projects}

The European Road Transport Telematics Implementation Co-ordination Organization (ERTICO) is an european ITS organization which have lot of projects in cooperative mobility, eco-mobility, safe mobility and info-mobility. COMeSafety2 (Communication for eSafety 2) is a part of projects on cooperative mobility (January 2011-December 2013), they work on standardization issues, best practices from European, Japanese and US field operational tests (FOTs) and an ITS architecture concept. COMeSafety architecture was used in the PREparation for DRIVing implementation and Evaluation of C2X communication technology (PRE-DRIVE C2X) project (2008-2010) which the main goal was to deploy simulation tools to estimate benefits in terms of safety, efficiency and environment. Then DRIVE C2X (DRIVing implementation and Evaluation of C2X communication technology, 2011-2013) its successor, includes lot of projects: DITCM (Ducth Integrated Testsite Cooperative mobility) in Neatherlands, SimTD (see section \ref{SIMTD}) in Germany, SCORE@F (see section \ref{SCORE@F}) in France with the project PRESERVE based on security on the vehicular communications, Coop TS Finland (Cooperative Test Site Finland), SAFER (Vehicle and Traffic Safety Center) in Sweden, SISCOGA (SIStemas COoperativos Galicia) in Spain and Test Site Italy in Italy. The tests are like SimTD, namely applications on traffic flow, traffic management, local danger alert, driving assistance, internet access and local information services, and some applications specific to each test field.

\paragraph{}SCANIA in collaboration with the Swedish National Road and Transport Research Institute tested the same use case as Volvo with SARTRE. Their test platoon with three-four trucks. Their questioning like all V2X projects is how to ensure the communication, who's responsible if something goes wrong, and who should pay for the technology.\\
Finally ConnectSafe is a field trial on road demo in Australia conduct by Cohda Wireless and the University of South Australia. They work on same use cases as the other V2V and V2I projects.

\paragraph{}A complete summary of all past and current V2V and V2I projects is in appendix of this report.
 
\subsection{V2U}
\paragraph{}In this section we will present innovative applications for the users. Lot of automobile automakers propose infotainment systems with a connection with web contents. We present also projects on the future smart car, most of them are not currently realistic but contain interesting ideas for include the user in the vehicular network.  

\subsubsection{Nokia: Connected Car Fund}
\marginnote{\includegraphics[scale = 0.3]{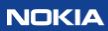}}

In 2014, Nokia launches a $\$100$ Million fund to invest in firms and start up specializing in connected and smart vehicles. Indeed they want to identify and invest in companies whose innovations will be important for a world of connected and intelligent vehicles, and they want to make investments that also support the growth of the ecosystem around their system of mapping (HERE) and their location products and services. They think that vehicles are becoming a new platform for technology adoption very similar to phones or tablets. They invest on a personal mobility.

\subsubsection{W3C launched work on Web Automotive}
\marginnote{\includegraphics[scale = 0.1]{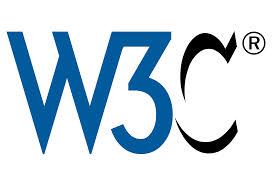}}

W3C (World Wide Web Consortium, ERCIM is the European host of W3C) has created an Automotive and Web Platform Business group to accelerate the adoption of Web technologies in the automotive industry. Developers, automotive manufacturers and suppliers, browser vendors, operators and others discuss how to enhance driving, safety and entertainment with a Web Open Platform. The group will first focus on defining a Vehicle Data API that will create new opportunities for automotive services via Web. The group Intel, QNX, Genivi (LG) and Webinos decided to focus on an API that would provide read access to vehicle data for the user. Next steps include creating an overview of the set of the proposed Vehicle APIs to look for overlaps and gaps, analyzing datasets shared by different OEMs via Vehicle API.

\subsubsection{Toyota: 'Toyota Friend'}
\marginpar{\includegraphics[scale = 0.2]{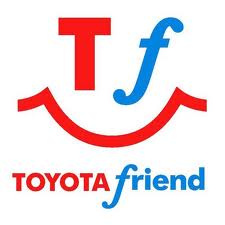}}

This is a partnership between Toyota and Salesforce \cite{ToyotaFriend}.
Salesforce is a global enterprise software company headquartered in San Fransisco. Their association started in may 2011 and first implementation on Toyota's electric vehicles and plug-in hybrids started in 2012 in Japan. This is a private social network which connects customer of electric vehicles, the dealership and Toyota. 
Members share maintenance tips like tweet with a service available on smartphone, tablet PCs or other mobile devices. Toyota offers a wide range of information on products and services on basic maintenance. Each driver can be notified by a 'tweet', for example when the battery from the electric or hybrid vehicle needs recharging. Moreover, Toyota Friend is a private social network, but customers can choose to extend their communications to family, friends and others on public social networks like Twitter or Facebook.
\begin{figure}[!htb]
\begin{center}
\includegraphics[scale = 0.3]{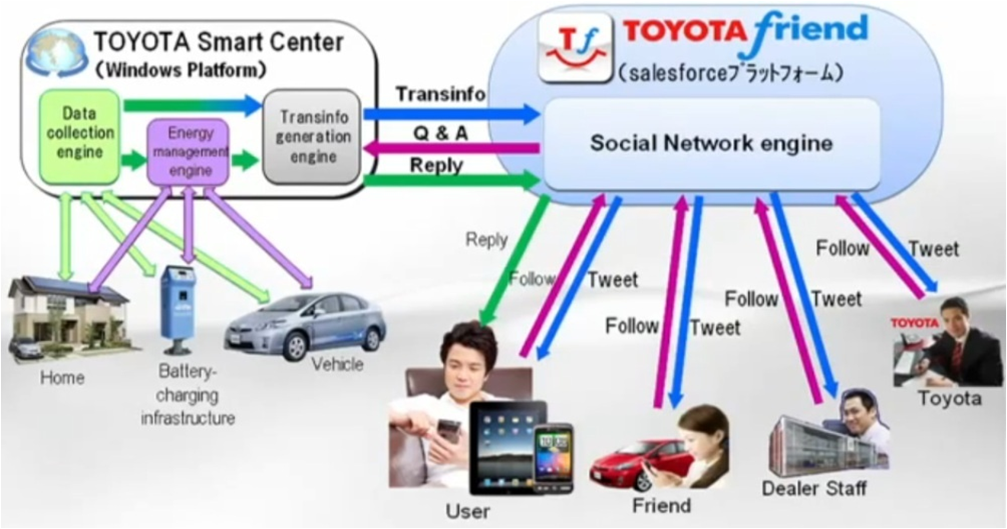}
\caption{Toyota Friend}
\end{center}
\end{figure}

\subsubsection{Alcatel-Lucent and Toyota: LTE connected car}
\marginnote{\includegraphics[scale = 0.1]{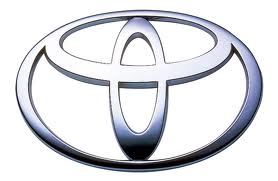}}
\marginnote{\includegraphics[scale = 0.05]{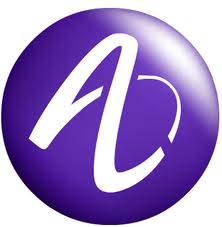}}

It was a concept car \cite{Alcatel} included in ng-connect program. The key outputs of ng-connect are prototype solutions that consist of expertise and technology from member companies (Alcatel-Lucent is a member). It was presented in 2009. The device is a touch screen which provides multiple applications to the driver and the passengers like driving assistant (in-vehicle parameter, navigation, maintenance, etc), web connection (music, video, games, etc), communication (phone call), internet of things in the vehicle through a connection with home called 'home control' with home security, light control, door locks, camera monitoring, energy management and multiple appliance control.
All applications are available thanks to the cloud computing with a distant access to a server.
\begin{figure}[!htb]
\begin{center}
\includegraphics[scale = 0.7]{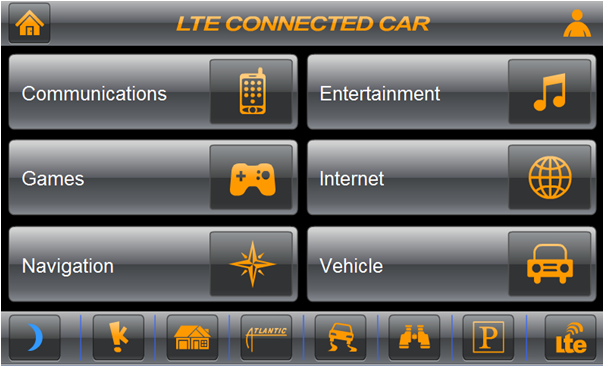}
\caption{Alcatel-Lucent and Toyota: LTE connected car}
\end{center}
\end{figure}

\subsubsection{Mercedes: 'Fleet tweet'}
\marginpar{\includegraphics[scale = 0.1]{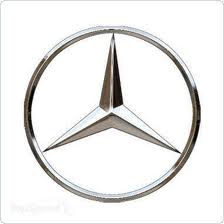}\\Fleet tweet}

This product \cite{Tweet} was marketed in December 2011. The system 'Park Assist' of Mercedes Benz (with a $360\,^{\circ}$ camera and four separated cameras) can recognize empty places by simply passing them. And when the car knew where the empty places are, everybody should know, Mercedes tweet empty parking spaces on the smartphone. The GPS system was connected to this Twitter account to tweet in real time all the precious space available based on your geolocation. 

\subsubsection{Ford: 'Facebook and Ford SYNC Hackathon'}
\marginnote{\includegraphics[scale = 0.1]{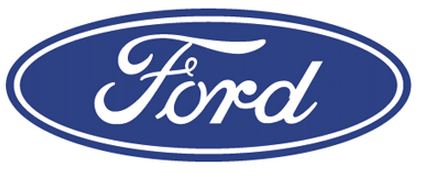}}
\marginnote{\includegraphics[scale = 0.04]{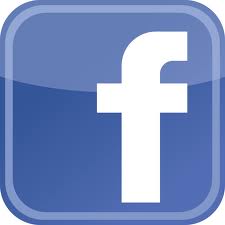}}

In February 2012 a Hackathon at Facebook's Palo Alto campus took place \cite{Facebook}. It was an experimentation on future social services in vehicle. They imagined and established services provide in the vehicle during only two days. The team integrates applications in car that allow driver to check-in somewhere as soon as he parks his car, find a restaurant that one of his friends 'likes', automatically stream songs that his friends have posted on their walls, post vehicle milestones, tag position, use Facebook messenger (system reads messages).

\subsubsection{Microsoft and West Cost Custom: 'project Detroit'}
\marginnote{\includegraphics[scale = 0.3]{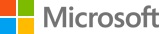}}

In a recent job posting, Microsoft Connected Car Team exposed their plans to leverage the full power of the current and upcoming Microsoft products and technologies in June 2012 \cite{West}. Project with Kinect in the car gave an overview in a demo vehicle in may 2012. It's a concept car, exactly it's an experiment to outfit a 1967 Ford Mustang Fastback with a suite of technologies: Kinect, Windows Phone, Windows 8 Tablette, WIFI in 4G, Windows Azure, HUD (Head Up Display). For Microsoft, the future connected car will know its drivers, and will interact with them naturally via speech, gestures, and face tracking. It will learn their habits, and offer personalized contextual information and driving assists to get them to their destination as quickly and safely as possible. Services allow to locate, lock, unlock and start the car remotely. A head up display has classic application like navigation, maps and turn by turn direction. It displays also points of interest: restaurants, gas stations, shopping centers. Driver can display message on the back windshield. Moreover system has a motion detector (front and back), camera that videos can be streamed on the phone (4G). Driver can use his windows phone like a microphone for the car's external audio systems. He also controls with his smartphone: the accent lighting, the horn sound, the activation of the projector screen and messages posted on the windshield. Finally driver and passengers can play at Xbox 360 on the back windshield.

\paragraph{}Microsoft partnered with some automakers create integrated telematics products (in-vehicle infotainment) such as Fiat Blue\&Me, Ford SYNC, Kia UVO. Kia UVO is car infotainment system with advanced voice and touch activated features. Ford SYNC is a factory installed, integrated in vehicle communications and entertainment system that allows users to make hands free telephone calls and control other functions using voice commands. And Fiat Blue\&Me allows to use mobile phone with Bluetooth technology and listen music thanks to the hands free function and steering wheel controls. Microsoft also delivered various iterations of Windows embedded automotive platform to the automotive suppliers such as Honda, Nissan, BMW, Mercedes-Benz, Aston Martin, and McLaren. Moreover Toyota and Windows work together on Toyota Media Service with the idea to connect hybrid and electric Toyota vehicles with a Windows Azure Cloud for in-vehicle infotainment.

\subsubsection{Mercedes: Mbrace2}
\marginnote{\includegraphics[scale = 0.15]{figures/LogoProjects/Mercedes.jpg}}
\marginnote{\includegraphics[scale = 0.4]{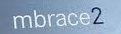}}

Mercedes Presented in 2012 at the CES (Consumers Electronics Show) their futur HMI which will be marketed in 2013: Mbrace2 \cite{Mbrace2}. 
It is an improvement of Mbrace. Mbrace represents an assistance system connected to a central unit and enables remote contact with the vehicle via smartphone or computer at any time. It offers safety services like directly report an accident, call emergency, notify the user if an attempt is made to break into his car or if it is stolen, find the car in a large parking, lock and unlock it remotely. Moreover traffic information and local weather are immediately available and many of the functions are directly accessible via the smartphone.

\paragraph{}In Mbrace2 users will have access to Facebook, Yelp and Google Local research for communicate his location with his friends, let them know when they arrive, search a restaurant, store or night place, purchase cinema tickets and make a restaurant reservation. An option of Geofencing is available, user set an area and if his car crosses this area a texto informs the owner. Mbrace2 enables a breadcrumb service which lets the owner to see where his car was driven. It connects cars with cloud in order to upgraded new software and applications: 'networked vehicle that is always online, always upgradeable'. Safety is not neglected, many functions are blocked while the car is driving and if there is a car problem an automatically alert both warns the user and his authorized Mercedes-Benz dealer.

\subsubsection{Bosch and University of St.Gallen: IoTS Lab}
\marginpar{\includegraphics[scale = 0.4]{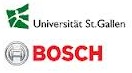}}

Bosch and the University of Gallen constructed a Laboratory in order to study the Internet Of Things (IoT) \cite{IoTS}. It started on March 2012 and first results are expected at the beginning of 2013. In this context Bosch develops a pilot project in Singapore called 'E-mobility', which will provide services for users of electric vehicles and propose the best way to reach their destination taking into account their electric vehicles' ranges, the charge spots available, the current traffic situation and the possibility to use public transportation. IoTS Lab works on Connected Mobility, Energy, City, Industry and Living. In the Connected Mobility, they are developing a business model that provides users with information on tourist attractions, leisure pursuits, and available charge spots. 

\subsubsection{Researchers: Project SiAM (DFKI)}
\marginpar{\includegraphics[scale = 0.7]{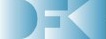}}

Three researchers \cite{DFKI} who work on the Project SiAM (situation adaptive multimodal interaction for innovative mobility concepts for the future) in the DFKI (German Research Centre for Artificial Intelligence) were awarded in the 'Intelligent Environments 2012' in August with a publication: 'Personalized In-Vehicle information Systems Building an application Infrastructure for smart Cars In smart Spaces'. The name of their application is \textit{the eye Box}. The application consists of a communication with the car on the environment around it. For instance you have just to say in your car: 'what is this building' or 'Give me more information on this monument' and the system answers thanks to the held of advanced eye tracking technology and algorithms. It is also able to recognize your gender and your approximate age. These researchers work also on SimTD (see section \ref{SIMTD})

\subsubsection{Toyota: 'Windows to the world'}
\marginpar{\includegraphics[scale = 0.1]{figures/LogoProjects/Toyota.jpg}\\Windows}

Toyota \cite{ToyotaWindows} works with Copenhagen Institute of Interaction Design (CIID) in 2011 in order to think about a concept around hobby for passengers. In their project they can draw image on the windows (visible outside), identify objects in another language, have the possibility to translate, view the distance of landmarks, zoom on the landscape and identify constellation in the sky. Currently two prototypes exist.

\subsubsection{General Motor: 'Windows of opportunity'}
\marginpar{\includegraphics[scale = 0.2]{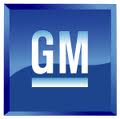}\\Windows}

A futurist project \cite{GMWindows} was given at a student's team of Bezalel academy of art and design in Israel by General Motors on the topic 'Windows of opportunity' (01/2012). The principle is to touch screen to distract the passengers with multiple applications: \\
Otto: information on landscape and weather,\\
Foofu: draw on the windows (like with the Mist),\\
Spindow: see the windows of any people in the world,\\
Pond: share music and message with other vehicles equipped.
\begin{figure}
\begin{center}
\includegraphics[scale = 0.5]{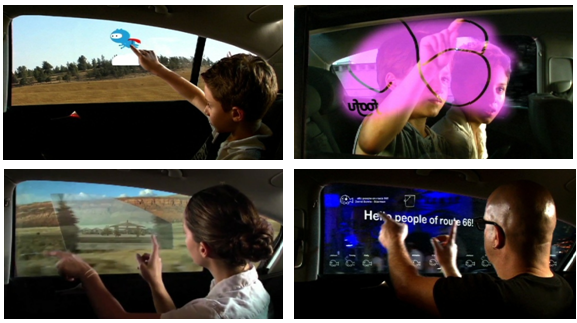}
\caption{Applications imagined by GM and students}
\end{center}
\end{figure}

\subsubsection{Audi: Future Urban Personal Mobility}
\marginpar{\includegraphics[scale = 0.25]{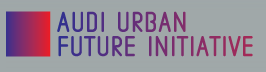}}

Audi and RMIT \cite{AudiUrbanMob} (Royal Melbourne Institute of Technology ) University's games and Experimental Entertainment Labo: GEELab (Created in 2011) work together on future urban, personal mobility, urban development, and the role of the Internet and technology in all this. They present a new urban flow services: games for rear seat with holographic 3D for children with gesture interactions on knowledge about the passing environment, flirting application between cars, rear seat can be like an office. They imagine also a time travel application to learn about current or past environment, a social application to find user's nearest friend and meet him. User can chose his type of driving behavior, a sporty behavior will cost credits. The system will display information on the car's exterior (battery information or your mood) and change its color. The main topic in all these future applications is the interaction with the environment. And the most interesting in their project is the fact that they view the city like a network with node and link not like streets and buildings.
\begin{figure}
\begin{center}
\includegraphics[scale = 0.4]{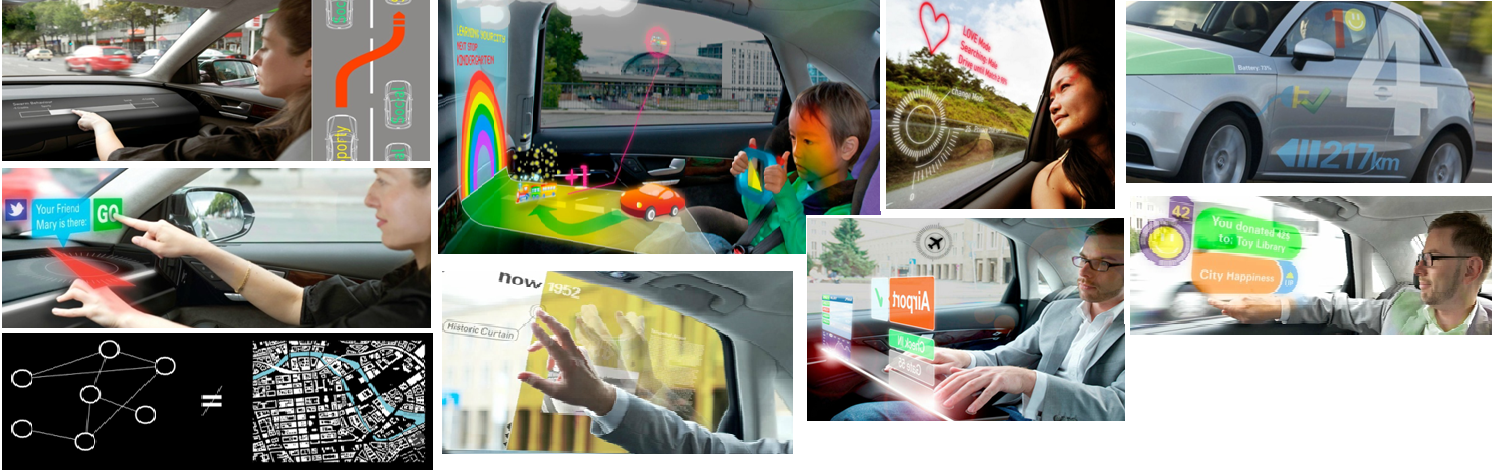}
\caption{Applications imagined by Audi and GEELab: the city is a network}
\end{center}
\end{figure}

\subsubsection{BMW: Connected Drive Concept Car}
\marginnote{\includegraphics[scale = 0.2]{figures/LogoProjects/BMW.jpg}\\Connected Drive}

BWM \cite{BMWConnected} present a concept car with a touch dashboard and a screen for passenger. Passenger can connect the car to the internet in order to search restaurants, make reservation, etc. Then he can send available information to the driver directly (swapping information). Moreover the system has a head up display for navigation, rear view camera and real time traffic information. BMW show an other innovation on their connected drive concept, user can send information of navigation from computer to the car.

\newpage

\subsubsection{Mercedes: 'DICE' (Dynamic Intuitive Control Experience)}
\marginnote{\includegraphics[scale = 0.15]{figures/LogoProjects/Mercedes.jpg}\\Dice}

Mercedes presented in 2012 at the CES (Consumers Electronics Show) in Las Vegas a concept car in research (it's planned for 2022) \cite{DICE}. Their idea is to introduce the augmented reality in the interior of car in order to present the content of social network and web by an interactive and dynamic way. The technology allows to control various aspects of the vehicle's infotainment system through simple hand movements (the system is like 'Kinect' of the Xbox360 by Microsoft). Many connected devices are reachable with fingers and eyes thanks to a head up display, a touch screen (the dashboard) and a gesture detection. An example of application is that driver can see the town map on the windshield with the places where are his friends thanks to the reports made by them on social network. Other services provide by DICE are: picture of friend displayed on the windshield when user is near his address and view the status updates for other simulated friends that have the DICE system. Points of interest are also related to the driver: restaurant, hotel, historic information on city (identifying buildings). User can receive in streaming and save music which is playing in a Bar or a Club when he pass in front of. Restaurant menu, map of the city, messages received and the current location of friends can be displayed. All these information can be read by the system. Functional aspects are mixed with social and emotional aspects. For example, when driver pass near a bar or a club it is possible with a simple gesture to play in streaming the music of this place.

\subsubsection{Toyota concept car (2011-2012), Diji = Fun Vii = iimo}
\marginnote{\includegraphics[scale = 0.2]{figures/LogoProjects/Toyota.jpg}\\    iimo}

This is a personalized connected car \cite{iimo} with a giant exterior screen capable of a recognition of the driver (security). User can change image via smartphone, tablet, etc (advertisement is also possible). He can see the localization of friends on the city map and check up of car before starting a trip (possibility to contact the dealership). 

\paragraph{}Inside the car all is a single screen (dashboard and car floor) that you can change. A menu in HUD (head up display) with temperature, mail, settings, weather, health, map, etc is displayed and user can easily update all services in the car. A virtual social life is available but all is a giant screen therefore users have to use the auto pilot. For the navigation system an hologram guide you near points of interest. Communication with other car is available for safety, to avoid accidents (maybe 802.11p) and also for playing (virtual game with friend in auto-pilot). 

\subsubsection{Visteon e-Bee Vehicle Concept User Experience Around the Car}
\marginnote{\includegraphics[scale = 0.2]{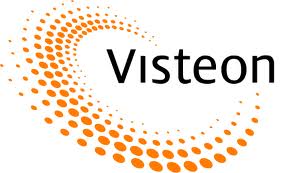}}

This project \cite{Visteon} was presented at the Electronica 2012 Trade Show in Munich (November 13-16 2012). An application allows users to share cars in the world with a remote personalization and a car remote control. An other application provides shopping inside the car through a detection of the advertisement in the environment (the delivery of the packet will be done in the car). There is also a driver detection which allows a personalized fragrance, an access to the agenda etc. An other service is around the safety with a detection of emergency vehicle. 

\subsubsection{Aeon Project}
\marginnote{\includegraphics[scale = 0.1]{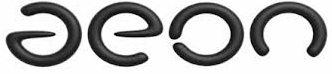}}

Aeon Project by Michael Harboun and Dassault Systems obtains the first prize at the 'Imagine Awards For Best Design and Communication 2011'. They imagine the futur car with several functions like automatic and manual drive modes, an interaction with social contacts, information about the region, the environment (tourism) which are displayed on a screen in 3D (augmented reality).

\subsubsection{Valeo Project}
\label{Valeo}
\marginnote{\includegraphics[scale = 0.1]{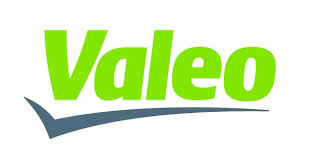}}

Valeo wants to investigate on the potential of the communication between cars and infrastructure through an efficient and innovative application for users. The applications chosen are the fluidity at an intersection and the green way (fluidity along a path). They take in consideration a real problem in the traffic: the congestion. The algorithms will not be centralized (like in the article \cite{algo}) but they will use the communication between cars in order to have a local optimization which will lead to a global optimization. Then the preferences of the user are taken into account in term of mobility in his car but also with his social network. In their project the dissemination of information take an important part in the vehicular network but also in the user network.

\subsubsection{Other Projects}
\paragraph{}
All these products will be marketed in 2014 \cite{logo}. 
\paragraph{}
\marginnote{\includegraphics[scale = 0.2]{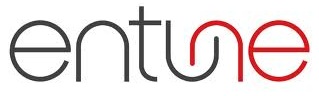}}
Toyota with Entune (touch screen, voice recognition, Bluetooth) includes Bing research for destination (with voice recognition) and phone calls. It checks traffic, sport scores, weather, gas station price and personal drive statistic. It also includes a personalized radio, a possibility of a reservation (restaurants, movies) and download music.
\paragraph{}
\marginnote{\includegraphics[scale = 0.15]{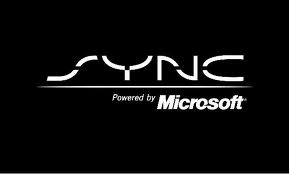}}
Ford with Sync (touch screen, voice recognition, Bluetooth USB, card SD, jack) includes navigation like GPS, can read message, controls in-vehicle climate function, has an application for phone with voice control (call, music search) and for station radio with voice control. It is able to transfer a current call on the car (uninterrupted connections), play music of user's phone, download contacts and photos, customize the screen, display Video DVD when vehicle is parked, provide an assistance, emit info on news, sport, weather and it has an internet connectivity 
\paragraph{}
\marginnote{\includegraphics[scale = 0.2]{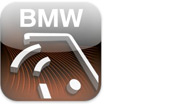}}
BMW has same options with iphone applications. 
\paragraph{}
\marginnote{\includegraphics[scale = 0.1]{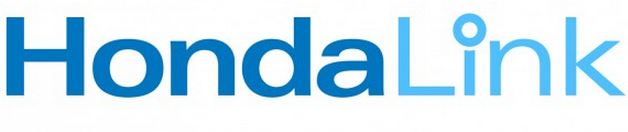}}
Honda with HondaLink has Bluetooth, USB, touch screen and Aha radio. It's able to speech text of Facebook, Twitter and e-mail, so user can listen but can't reply. With Yelp it offers Restaurant and coffee recommendation, a personalized web radio, a voice to text SMS services and its own smartphone application.

\section{Summary and discussion}

\begin{figure}[!htb]
\begin{center}
\includegraphics[scale = 0.4]{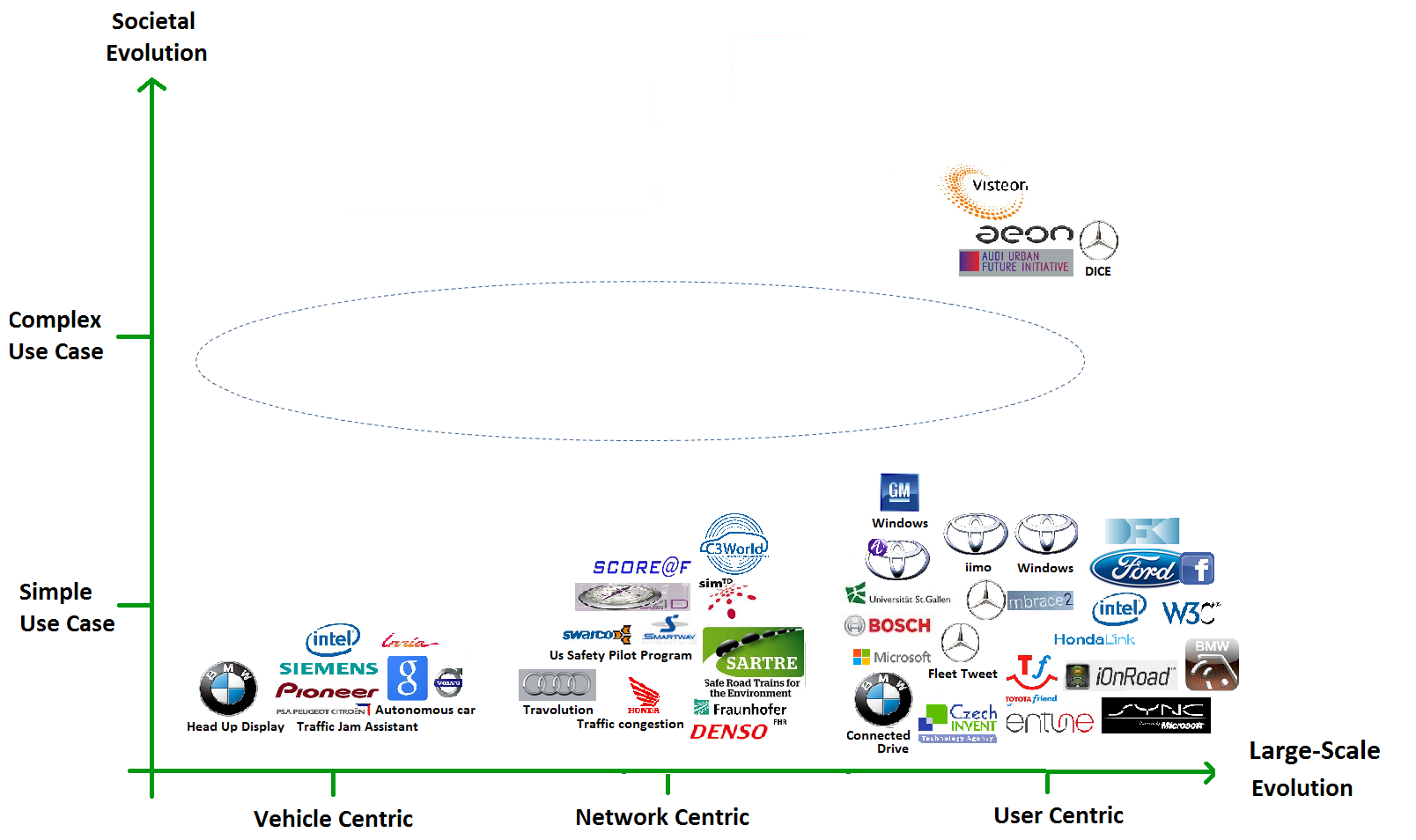}
\caption{Existing projects according to their use case complexity}
\label{classification}
\end{center}
\end{figure}

\paragraph{}Following our classification presented in section \ref{Classification}, we are able to classify all these projects according to their societal evolution and scalable evolution (see figure \ref{classification}). Some projects are exclusively focusing on the vehicle without the consideration of the driver or passenger or other vehicles, as the communication with traffic light in Travolution project by Audi or the system of navigation assistant with head up display by BMW, Pioneer, Intel or Siemens. Another example is Peugeot and Citro\"en, which proposed a relevant research of information on the web, but exclusively for the vehicle (parking, navigation, etc.). They consider V2RL bricks and so only lay in the vehicle-centric view (see figure \ref{VehicleCentric}). Then other projects like field operational tests called SimTD for Germany, or the Connected Vehicle Safety Pilot Program for United States, concentrate their studies around the connection between the vehicle and the infrastructures, belonging to the network-view. Finally, most projects consider the driver and passengers with a connection with their web contents (Sync, Entune, HondaLink, etc.) or the environment (DFKI). But, in these three cases, on the horizontal axis, most of these projects can only be represented as simple use cases and a gap between the vehicle and the user still exists. Through these applications, the user and the vehicle are separated entities, only considered in different contexts. Nowadays, the smartphone is fully an extension of the user, and the future projects like Aeon or Visteon adopt this view for a key deployment of the smart car.

\paragraph{}Some projects, finished or not, only catch the environment for the safety of the vehicle, considering only V2RL blocks (Vehicle to Real World, see figure \ref{VehicleCentric}). Here, we do not have a connection with the environment or other vehicles, it is just a view of the environment by the car. For instance, Pioneer, Intel, BMW Head-up-Display, Siemens use simple bricks (speed detection, sign detection, safety distance or lane detection); iOnRoad belongs to this category with a brick on safety distance but it allows an original connection to the user with simple features hence its position near the user-centric view in our classification \ref{classification}. Autonomous car by Google and traffic jam assist by many car makers (BMW, Ford, Volkswagen, GM, Audi) catch only the environment with a camera, a radar, etc. PSA Peugeot Citro\"en show relevant web information for navigation. There are no communication V2V or V2I, so we consider that it is a vehicle-centric view with simple bricks of detection.

\paragraph{}Then all field operational tests on V2V and V2I work on basic use case (see figure \ref{VehicleCentric}) in order to show the utility of the communication between cars on safety and traffic efficiency like SimTD, Score@f, ConnectSafe, Compass4D, Swarco and the US Safety Pilot. Honda in its traffic congestion project uses distance and speed coordination, SARTRE uses the same tools of communication with infrastructure to manage the process, Audi with Travolution interacts with traffic lights, Denso also uses this communication and detection of emergency vehicle, SmartWay shows ideas around V2V. C3World is slightly more complete but focus on research topics. Consequently they belong all to the network-centric view with simple use cases.

\paragraph{}Finally, many projects are concentrated on the user's environment like ToyotaFriend which connects the user with his concession or Alcatel which connects the driver with his home. Project Detroit of Microsoft (message on the windshield), Bosch IOT (best way between two points with the consideration of the energy consumption, tourism), DFKI (tourism), Toyota windows, and GM windows also allow a connection with the environment for passengers or drivers. Moreover, General Motor windows can be classify near the network-centric view because its application allows a connection with other vehicles (sharing of windows). All these projects begin to create a contact between the user, the vehicle and the environment, but most of them will only be possible and available in many years.

\paragraph{}With the advent of web identity and online social network, car makers see an important interest to include web contents to the drivers through a basic way. For example, TweetFleet by Mercedes uses Twitter to communicate empty parking places and Ford introduces content of Facebook in the car. Mbrace2 by Mercedes, Connected Drive and Iphone application by BMW, Entune by Toyota, Sync by Ford, HondaLink and Intel show web pertinent information for the user (restaurant, weather, etc.) or/and safety services or/and personal and adapted services to the driver and passengers. These services are provided with a strong relationship with the smartphone and technologies like gesture detection, 3G, Bluetooth, etc. Finally the most elaborated projects which take into account the user with the environment and his social identity are futurists like Audi urban Future, DICE, iimo, Visteon or Aeon.

\paragraph{}Through the diagram (in figure 20) we notice a global tendency to take into account the profile of the driver in his car. However - except to futurist projects -, car makers stay around a vehicle-centric view, network-centric view or very basic user-centric view with simple use case. So fort, no correlation exist between these three views. A gap really exists between the user network and the vehicular network. To answer to the future needs of the user in his connected car, applications and services will have to connect user social environment with the vehicle network. A connection of the bricks in all views (see figure \ref{VehicleCentric}) will lead to this kind of services and will give a complete use case that will transform a simple car in a smart car. 


\section{Conclusion}

\begin{figure}[!htb]
\begin{center}
\includegraphics[scale = 0.5]{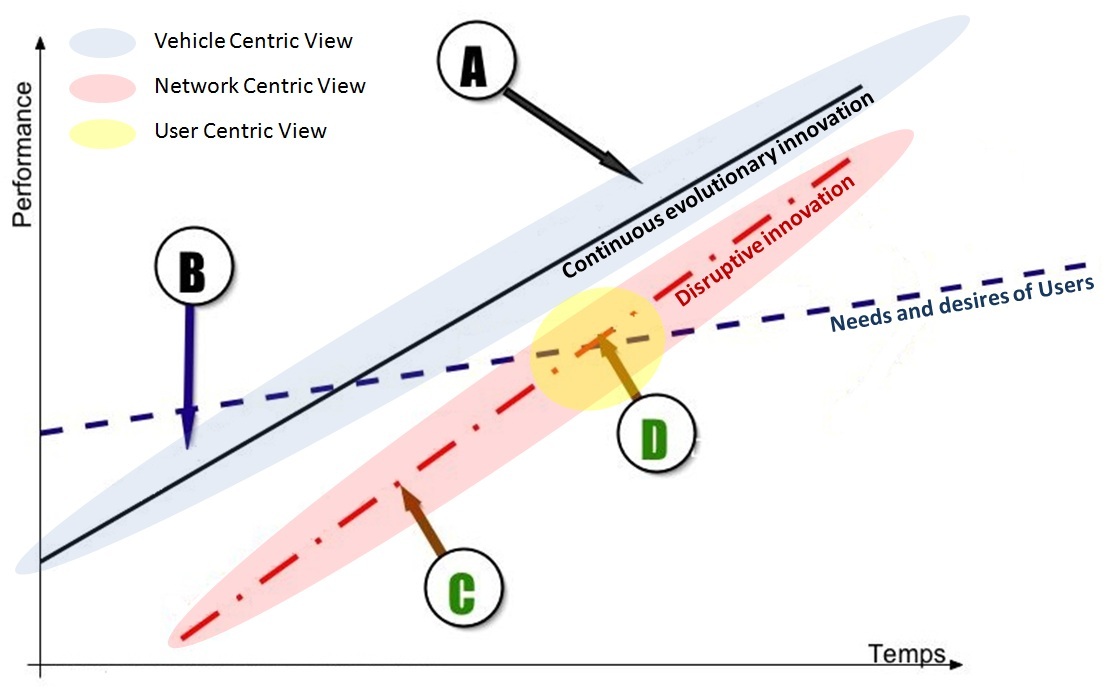}
\caption{Vehicular innovation breakthrough points}
\label{Innovation}
\end{center}
\end{figure}


\paragraph{}To conclude this survey, we illustrate the innovation process in the vehicular context. The following figure - inspired by research of the professor Clayton M. Christensen \cite{Christensen1997} - summarizes the context of innovation according to our study. 

\paragraph{}The B point represents existing products - with few performance - that have raised from the beginning of the in-vehicle information technologies. In this case, the vehicle is still only a transportation facility and only need surety and security improvements. As vehicles are also in improvement for comfort and entertainment, the obvious evolution from B point to A point is the addition of options and accessories around the vehicle. In A point, innovations are above the needs of users (and often expensive). This line of continuous evolutionary innovations is currently widely explored, focused on the vehicle, but tend to stagnate. 

\paragraph{}With the increasing number of vehicles and the great development of smart cities and urban/societal concerns, new issues about traffic jam and congestion have emerged. The communication between cars and infrastructure can be an answer and - at the same time - bring innovations in the context of security, comfort and entertainment. The C point illustrates this disruptive innovation and represents the new market. Only few early products are present now on this market, but the interesting, innovative research issues and the user satisfaction will meet at the D point where users, vehicles and the environment will profit and interconnect.

\appendix


\section*{Supplementary material}

| PROJECTS | | | | | | |
|---|---|---|---|---|---|---|
| NAMES | LOGOS | DESCRIPTION | ZONE | PARTICIPANT | SITUATION | LINK |
| COMeSafety | 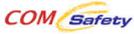 | The overall goal of COMeSafety is to support realisation and possible deployment of cooperative communication-based active safety systems | EU | BMW (project coordinator), Daimler, Fiat, Renault & Volvo | FINISH (12/2013) | www.comesafety2.org |
| Co Cities | 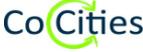 | pilot project to introduce & validate cooperative mobility services in cities & urban areas | EU | | FINISH (12/2013) | www.co-cities.eu |
| COSMO | 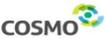 | instal & run practical demonstrations of a range of V2V &V2I services in realistic conditions | EU | Volvo, Fiat | FINISH (06/2013) | www.cosmo-project.eu |
| SmartCEM (Smart Connected Electro Mobility) | | project is designed to demonstrate how ICT solutions can make commuting in electric vehicles more practical and overcome shortcomings associated with them (smartCEM 2012) | | | | |
| Drive C2X | 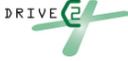 | equipped with radio hardware based on 802,11p & UMTS for data exchange with other vehicles or roadside infrastructure | EU | (coordinated by Daimler), Audi, BMW, Fiat, Ford, Honda, Opel, PSA, Renault & Volvo,Bosh, Continental, Delphi, Denso, Hitachi, NEC & Renesas,Traffic engineers: PTV (Planung Transport Verkehr) ERTICO ITS, EICT | FINISH (12/2013) | www.drive-c2x.eu/overview |
| eFrame | | | EU | | FINISH (05/2011) | |
| EuroFot | 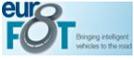 | testing 8 ADAS functions to assist the driver in detecting hazards & avoiding accidents | EU | each funtions is tested by differents OEM as Ford, MAN, Volvo cars, Volvo Trucks, VW & Audi, Renault, Fiat, [BMW, Daimler (Safe HMI)]Bosh, Continental, Delphi & HarmanAlcor, ERTICO, ADAS, EICT | | www.eurofot-ip.eu |
| FOT-Net | 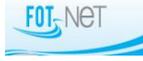 | large-scale test programmes aiming at a comprehensive assessment of the efficiency, quality, robustness and acceptance of ICT solutions used for smarter, safer and cleaner and more comfortable transport solutions. | EU | Daimler AG, Ford, Nissan,Orange, Continental, Vufo GmbH,VTT (coordinator), ERTICO-ITS Europe, Chalmers University of Technology, IKA, CTAG, University of Leeds, | (01/2014 duration of 36 months) | www.fot-net.eu |
| FOT-sis | | testing of the road infrastructure management systems | EU | | | www.fotsis.com |
| GeoNet | | The goal of GeoNet was thus to implement and formally test a networking mechanism as a standalone software module which could be incorporated into Cooperative Systems. | EU | | FINISH (01/2010) | www.geonet-project.eu |

| Name | Logo | Description | Region | Partners | Status | Website |
|---|---|---|---|---|---|---|
| Intersafe-2 | 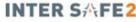 | it aims to develop & demonstrate a cooperative Intersection Safety System (CISS) that is able to reduce injury and fatal accidents at intersections. Diminution of 80% of accidents & benefit of up to 40% | EU | BMW, VW | FINISH (2011) | www.intersafe-2.com |
| OVERSEE | 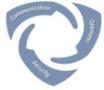 | it will ensure security, reliability, secure communications & strong mutual isolation of simultaneously running applications | EU | VW,Industrial Partners (escrypt GmbH - Embedded Security TRIALOG OpenTechEDV Research GmbH) Academic Partners (Technische Universität Berlin Fraunhofer Universidad Politécnica de Valencia | FINISH (06/2012) | www.oversee-project.com |
| Pre-Drive C2X | 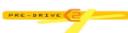 | | EU | Volvo, Daimler | FINISH (06/2010) | |
| TeleFOT | 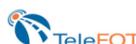 | most important project CVIS managed by european ITS organization ERTICO | EU | BMW, Daimler, Renault, Volvo, Fiat, Magneti,Bosh,Navteq, Tele Atlas, Telecom Italy, Vodafone | FINISH (05/2012) | www.telefot.eu |
| IntelliDrive | 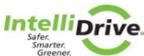 | formerly Vehicle Infrastructure Integration. Focused on developping standardized wireless vehicular communications. The US DOT developped and test a 5,9GHZ-based VII POC | USA | | FINISH (2009) | |
| SmartWay | 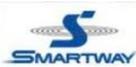 | next generation system that incorporates existing services such as VICS & Electronic Toll Collection System. All major auto, electronics & governement organizations are involved. Japan are in advance compare to USA and EU due to his V2I based-system already deployed | Japan | all auto OEMs and their automotive suppliers | | |
| Safety Pilot Test | | | USA | GM, Ford, Toyota, Honda, VW, Daimler, Hyundai, Nissan,In-vehicle: AutoTalks, Cohda, Denso, Delphi, Visteon, DGE, ITRI, Savari, Arada,Roadside: Arada, Kapsch, ITRI,Cohda/Cisco Aftermarket Safety Devices: Kapsch | | |
| SIM TD | 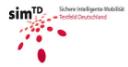 | The sim$^{TD}$ research project is shaping tomorrow's safe and intelligent mobility through researching and testing car-to-x communication and its applications | GERMANY | Opel, Audi, BMW, Daimler, Ford, VW,Bosch, Continental, Deutsche Telekom,NAVTEQ cambrionix AXIO-NET | | |

| | | | | | | |
|---|---|---|---|---|---|---|
| DSSS | | Driving Safety Support System It's managed by the UMTS of Japan with support from the National Police Agency (NPA). It's a V2I solution for ordinary roads to prevent traffic accident | Japan | | | |
| CVIS | 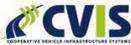 | focused on developping key elements needed to test and prove the viability of V2V &V2I systems | EU | Fiat, Daimler, Renault, Volvo,Bosch, Siemens, Vodafone,Autoroutes du Sud de la France, BAE Systems, CIT, Department of Transport, Efkon, ERTICO, FEHRL, INRIA, Intempora, Navteq | FINISH | http://www.cvisproject.org/ |
| FOAM | | it defines an architecture that connects the in-vehicle systems, roadside infrastructure and back-end infrastructure that is necessary for cooperative transpoort management. | EU | | | |
| iCar support | | iCarSupport targets at attaching public awareness to the perspectives opened by ITS technologies and applications in making vehicles smarter and safer. | EU | | | www.icarsupport.eu/ |
| iTETRIS | 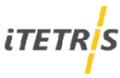 | The EU-FP7 project iTETRIS (An Integrated Wireless and Traffic Platform for Real-Time Road Traffic Management Solutions) develops an evaluation platform for large-scale, long-term simulations of cooperative traffic management applications. The platform combines traffic and communications simulation in a closed-loop which allows detailed analyses of effects and performance of cooperative applications on traffic flow, travel time, emissions, etc. | EU | Hitachi,Thales, CBT, Innovalia, PEEK, Eurecom | | www.ict-itetris.eu; http://www.eurecom.fr/en/publication/2785/download/rs-publi-2785.pdf |
| PRESERVE | 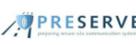 | "Preparing Secure Vehicle-to-X Communication Systems" The goal of the EU-FP7 project PRESERVE is to bring secure and privacy-protected V2X communication closer to reality by providing and field testing a security and privacy subsystem for V2X systems. | EU | Renault, Audi, BMW, Daimler, Volkswagen,Denso, Infineon, CAMP Consortium and simTD,University of Twente, Escrypt, Fraunhofer Institute for Secure Information Technology — SIT + AISEP, Trialog | | |
| EVITA | 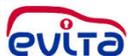 | The objectives of the EVITA project were to design, to verify, and to prototype building blocks for automotive on-board networks where security-relevant components are protected against tampering and sensitive data are protected against compromise. | EU | BMW,Bosh, Continental, Siemens,Fraunhofer SIT-ISI, escrypt, EURECOM, Infineon, MIRA, Trialog, Fujitsu | FINISH | www.evita-project.org |

| Name | Logo | Description | Region | Partners | Status | URL |
|---|---|---|---|---|---|---|
| PREVENT | 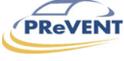 | "PReVENTive and Active Safety Applications" PReVENT aimed at developing, testing and evaluating safety related applications, using advanced sensor and communication devices integrated into on-board systems for driver assistance. | EU | DaimlerChrysler AG, BMW F&T GmbH, Centro Ricerche FIAT SpA, Ford, PSA, SAGEM Défense Securité, Volvo Technology AB and INRETS, Audi AG, Renault, Volkswagen **Research institutions or centres:** CERTH-HIT, CIDAUT, CNRS, Fraunhofer-IVI, FORWISS, HTW-FORGIS, IKA-RWTH,INRETS, INRIA, ICCS, LCPC, Lunds Universitet, NTUA, TNO, Technische Universität Chemnitz, Universita degli Studi di Parma, Universita degli Studi di Trento, Universität Hannover, Transver GmbH, VTT,Bosch, Delphi Delco Electronics, Siemens, Blaupunkt,FCS Simulator System, IBEO, Imita AB, Lewicki Microelectronic GmbH, Navigon AG, Navteq BV, Philips, Signalbau Hüber, Tele Atlas NV, TRL Ltd, TRW Ltd, | FINISH | http://en.wikipedia.org/wiki/PReVENT |
| SAFESPOT | 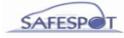 | "Smart Vehicles on Smart Roads". Cooperative systems using communication between vehicles and in the infrastructure via can considerably enhance this field of view, thus leading to a breakthrough for road safety. | EU | Fiat, Daimler, Renault, Volvo,Bosh, Siemens, Magneti, Continental,ANAS, Cofiroute, Mizar, Piaggio, IBEO, Kapsch, Lacroix, Navteq, PTV, G-Free, Télé Atlas, PEEK, ERTICO, CNRS, | FINSH | http://www.safespot-eu.org/ |
| SEVECOM | 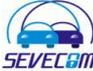 | "SEcure VEhicle COmmunication" Sevecom is an EU-funded project that focused on providing a full definition and implementation of security requirements for vehicular communications. | EU | Fiat, Daimler,TRIALOG Budapest University of Technology and Economics EPFL - Ecole Polytechnique Fédérale de Lausanne Katholieke Universiteit Leuven Ulm University,Bosch | FINSH | www.sevecom.org |
| AMAS | | Autonomus Mobility Applique System. low-cost sensors and control systems on **military vehicles** to enable driver assistance features or automated operation | USA | | | |

| Name | Logo | Description | Country | Partners | Status | Link |
|---|---|---|---|---|---|---|
| **Arizona E-VII Program** | | consisted of two projects under Arizona DOT: SPR-653, Arizona VII Initiative: Proof of Concept/Operational Testing and SPR-678, Dynamic Routing for Incident Management. Prototype applications for the program included traffic signal preemption and priority, ramp meter preemption, and mobile incident warning | USA | | | |
| **Mn Pass** | | designed to charge a fee for faster travel (less congested lanes), without the need to designate the entire road as a toll road. | USA | | | |
| **Autonomous Intersection Management** | 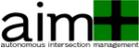 | how intersection control mechanisms can use autonomous vehicles in order to improve both safety and efficiency. | USA | | | |
| **Start ITS from Kanagawa, Yokohama (SKY)** | | Project goals were to ease traffic congestion and reduce accidents | Japan | Nissan Motor Co., Ltd., NTT DoCoMo, Inc., Panasonic Corporation, and Xanavi Informatics Corporation (now Clarion Ltd.) | | |
| **Instant Mobility project** | 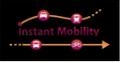 | centered on providing Internet access for transport and mobility | EU | | | |
| **Cooperative Mobility Pilot on Safety and Sustainability Services for Deployment** | | focuses on improving safety, energy efficiency, and congestion. | | | | |
| **AKTIV (Adaptive and Cooperative Technologies for Intelligent Traffic)** | 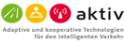 | backed by a consortium of 29 partners, developed an assistance system under its Cooperative Cars (CoCar) project. The goal of the initiative is to prevent accidents using intelligent traffic management systems and mobile communications technologies for connected vehicles | Germany | Opel, AUDI, BMW, Daimler AG, MAN, Volkswagen, Ford, Continental, Bosch, Siemens, Ericsson, Ibéo, Allianz, TeleAtlas, TRANSVER, Vodafone | | http://www.aktiv-online.org/english/projects.html |
| **HAVEit (Highly Automated Vehicles for Intelligent Transport)** | 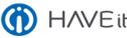 | project concentrated on partially automated vehicles explored how drivers interact with vehicles with different levels of automation | | Volvo, VW, Continental, DLR, Hadlex | FINISH (June 2011) | |

| Project | Logo | Description | Country | Partners | Status | Link |
|---|---|---|---|---|---|---|
| KONVOI (a German acronym for Development and Analysis of Electronically Coupled Truck Platoons) project | | was focused on the use of Advanced Driver Assistance Systems (ADAS) to form truck platoons of up to four vehicles on public roads that could improve traffic flow, fuel consumption, and environmental performance of heavy-duty highway vehicles. | Germany | | | |
| SCORE @F) Système Coopératif Routier Expérimental Français) | 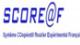 | goals for the SCORE@F project are to quantify benefits of the system, identify stakeholders, validate or evolve standards and applications, develop qualification tests to ensure interoperability, and calculate deployment costs | France | Renault, PSA, Cofiroute, Néavia, Senda, Viveris, Orange, Intempora, EURECOM | | http://www.pole-moveo.org/pdf-projets-das/Scoref-A.pdf |
| Automatisation Basse Vitesse (ABV) | 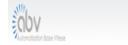 | was focused on automation for low-speed vehicles | France | PSA, INRETS, Continental, IBISC, IEF, Induct, INRIA, LAMIH, Viametris, UHA – MIPS, and Véolia Environnement Recherche & Innovation | FINISH (October 2012) | |
| Connected Cruise Control (CCC) | | project will result in a built-in solution to provide driving advice regarding speed, headway, and lane so drivers can anticipate and prevent congestion. The technology integrates invehicle and roadside systems to improve traffic flow | | NAVTEQ, NXP, SAM, Technolution, TNO | | |
| SARTRE (Safe Road Train for the Environment) | 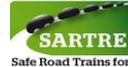 | The main goal of the project is to develop and test vehicles that can autonomously drive in long convoys or road trains | UE | led by Volvo, Tecnalia, Idiada, IKA, SP Technical Research Intistute of Sweden | | http://en.wikipedia.org/wiki/Safe_Road_Trains_for_the_Environment; http://www.sartre-project.eu/en/Sidor/default.aspx |
| Cooperative Systems for Intelligent Road Safety (COOPERS) | | project used existing equipment and infrastructure as a foundation when developing standardized wireless bidirectional infrastructurevehicle technology | UE | BMW, Efkon, Fraunhofer, Navteq, TFH | | |



\bibliographystyle{plain}
\bibliography{biblio}

\end{document}